\documentclass[aps,prd,amssymb,eqsecnum,showpacs] {revtex4}
\usepackage{graphicx}
\usepackage[mathscr]{eucal}
\usepackage{mathrsfs}

\begin{document}

\title{Testing the well-posedness of characteristic evolution of scalar waves}

\author{M.~C. Babiuc${}^1$, H-O. Kreiss${}^{2,3}$   and J. Winicour${}^{3,4}$
       }
\affiliation{
${}^{1}$Department of Physics \\
 Marshall University, Huntington, WV 25755, USA \\
${}^{2}$NADA, Royal Institute of Technology, 10044 Stockholm, Sweden\\
${}^{3}$Max-Planck-Institut f\" ur
         Gravitationsphysik, Albert-Einstein-Institut, 
	  14476 Golm, Germany\\
${}^{4}$Department of Physics and Astronomy \\
         University of Pittsburgh, Pittsburgh, PA 15260, USA 
	 }

\begin{abstract}

Recent results have revealed a critical way in which lower order terms affect
the well-posedness of the characteristic initial value problem for the scalar
wave equation. The proper choice of such terms can make the Cauchy problem for
scalar waves well posed even on a background spacetime with closed lightlike
curves. These results provide new guidance for developing stable characteristic
evolution algorithms. In this regard, we present here the finite difference
version of these recent results and implement them in a stable evolution code.
We describe test results which validate the code and exhibit some of the
interesting features due to the  lower order terms.

\end{abstract}


\maketitle

\section{Introduction}

We develop and test computational evolution algorithms based upon a recent proof
of the well-posedness of characteristic initial value and boundary problems for
a scalar wave~\cite{wpsw}. Well-posedness requires the existence of a unique
solution which depends continuously upon the data with respect to an appropriate
norm. Characteristic problems consist of finding a solution  of a hyperbolic
system of partial differential equations where the initial data is given on a
characteristic hypersurface, i.e. a lightlike hypersurface for the case of scalar
waves in Minkowski space.  Whereas, the Cauchy problem for initial data on a
spacelike surface is well posed with respect to an $L_2$ norm, this is not
necessarily true for the Cauchy problem for initial data on a characteristic
surface.   For example, the characteristic Cauchy problem for the equation,
$\partial_t \partial_x \Phi =S$, on the domain $t\ge 0, -\infty <x < \infty$
with compact source $S(t,x)$ and compact initial data $\Phi (0,x)$ on the
characteristic $t=0$, allows the solution freedom $\Phi(t,x)\rightarrow
\Phi(t,x)+ g(t)$, where $g(t)$ is independent of the initial data. This lack of
uniqueness caused by waves entering from past null infinity ($x=-\infty$) cannot
be remedied by requiring a finite $L_2$ norm (integrated over $x$) because this
would also rule out the existence of bonafide waves generated by the source $S$
which travel to $x=+\infty$ . However, the recent work~\cite{wpsw} showed that
the corresponding Cauchy  problem for the modified wave equation
\begin{equation}
    \partial_t (\partial_x +a) \Phi =0, \quad a>0
\label{eq:1dm}
\end{equation}
is well posed with respect to an $L_2$ norm. When translated back, this result
corresponds to the well-posedness of the original problem with $a=0$ in terms of
a finite $L_2$ norm that is exponentially weighted by the factor $e^{-ax}$,
which rules out waves entering from $x=-\infty$ but allows waves from the source
to propagate to $x=+\infty$. In this sense, the original problem is well posed
as an initial-boundary value problem with a  boundary condition at $x=-\infty$.
However, the benevolent behavior of the modified equation is more general. The
Cauchy problem for (\ref{eq:1dm}) on the periodic domain $0\le x\le 1$ is well
posed, and this property extends to higher dimensional wave equations, e.g. 
\begin{equation}
    \partial_t (\partial_x +a) \Phi +\partial_y^2 \Phi =0, \quad a>0 , \quad -\infty<y<\infty.
    \label{eq:aterm}
\end{equation}
This is a surprising result since the problem with $a>0$ is well posed even when
the characteristics in the $x$-direction form closed lightlike curves, in which
case a signal propagates instantaneously back to the emitter. The corresponding
problem with $a\le 0$ is ill posed. The importance of the condition $a>0$ on a
lower order term had not been previously recognized in the characteristic
initial value problem for the wave equation.

These results based upon the modified wave equation generalize to characteristic
initial value problems for quasilinear waves in higher dimensions. Furthermore,
this approach allows application to characteristic problems the standard
technique~\cite{klor} of proving well-posedness by
splitting the problem into a Cauchy problem and a local half-plane problem,  and
then proving that these individual problems are well posed. In particular, the
main result in~\cite{wpsw} is the well-posedness of the null-timelike
initial-boundary value problem for quasilinear scalar waves
\begin{equation}
        g^{ab}\nabla_a \nabla_b \Phi = S (\Phi,\partial_c \Phi, x^c)
        \label{eq:qw}
\end{equation}
propagating on an asymptotically flat curved space background with source $S$,
where the metric $g^{ab}$ and its associated covariant  derivative $\nabla_a$
are explicitly prescribed functions of $(\Phi,x^c)$. In coordinates based upon
the retarded  time $t=const$ outgoing lightcones, the metric takes the
Bondi-Sachs form~\cite{bondi,sachs}
\begin{equation}
   g_{ab}dx^a dx^b = -(e^{2\beta}W-r^{-2}h_{AB}W^A W^B)dt^2 -2e^{2\beta}dtdr
      -2h_{AB}W^B dtdx^A   +r^2h_{AB}dx^A dx^B,
      \label{eq:nullmet}
\end{equation}
where $x^A$ are angular coordinates along the outgoing light rays, $r$ is a
surface area coordinate and the coefficients $\beta$, $W$, $W^A$ and $h_{AB}$ 
have the appropriate falloff for asymptotic flatness. The quasilinear wave
equation (\ref{eq:qw}) with asymptotically flat metric (\ref{eq:nullmet})
becomes:
\begin{eqnarray}
     \label{eq:3dwe}
  {1\over r} (-2  \partial_t  \partial_r +W \partial_r^2) (r\Phi)
    +(\partial_r  W)\partial_r\Phi
    & -&\frac{1}{r^2}D_A(W^A \partial_r \Phi) 
        -{1\over r^2} \partial_r (W^A D_A \Phi) 
     +{1\over r^2} D_A(e^{2\beta} D^A \Phi) \nonumber \\
    & = &e^{2\beta} S(\Phi,\partial_c \Phi,x^c) ,
 \end{eqnarray}    
where $D_A$ is the covariant derivative with respect to the conformal 2-metric
$h_{AB}$. The null-timelike problem on this background consists of determining
$\Phi$ in the region $(r>R,t>0)$ given data $\Phi(0,R,x^A)$ on the initial null
hypersurface $t=0$ and on the timelike worldtube $r=R$. The well-posedness of
this problem was established in~\cite{wpsw} by considering the modified
problem in terms of the variable $\Psi = e^{aR/r} \Phi$, with $a>0$, which has the
same effect as the $a$-term in (\ref{eq:aterm}).

The use of null coordinates, as introduced by Bondi~\cite{bondi}, was key to the
understanding and the geometric treatment of gravitational waves in the full
nonlinear context of general relativity~\cite{sachs,penrose}. The physical
picture underlying the null-timelike problem for Einstein's equations~\cite{tam}
is that the worldtube data represent the outgoing gravitational radiation
emanating from interior matter sources, while ingoing radiation incident on the
system is represented by the initial null data. This problem has been developed
into a Cauchy-characteristic matching scheme in which the worldtube data is
supplied by a Cauchy evolution of the interior sources (see~\cite{lr} for a
review). Although characteristic evolution codes have been  successful at
simulating many such problems, the well-posedness of the null-timelike problem
for the Einstein equations has not yet been established. The recent proof of
well-posedness of the corresponding problem for the quasilinear wave equation
(\ref{eq:3dwe}) is a first step toward treating the gravitational
case~\cite{wpsw}.

In this paper, we show how this new approach can be implemented as numerically
stable characteristic evolution codes. The fundamental understanding of
characteristic evolution is in a primitive state in comparison to Cauchy
evolution. For that reason, it is most useful to restrict our attention to model
problems in a 3-dimensional spacetime. The stability of variable coefficients
and quasilinear problems is based upon the stability of constant coefficients
problems under lower order perturbations. The model problems considered here
provide the basis for the stable treatment of more general problems. In
Sec.~\ref{sec:alg} we establish the stability of characteristic evolution
algorithms for the Cauchy problem.  Test results for this problem are presented
in  Sec.~\ref{sec:testper}. Besides confirming the expected stability
properties, the simulations reveal interesting features.

In Sec.~\ref{sec:alghalf} we describe a boundary algorithm for the half-plane
problem and prove that it is numerically stable provided $a>0$. In
Sec.~\ref{sec:testbound} we test this algorithm, and simulate the effect of
boundaries. 

The standard finite difference conventions, notation and techniques are described
in~\cite{kgust}, where further details can be found.

\section{Evolution algorithms for the Cauchy problem}
\label{sec:alg}

We consider the Cauchy problems for a scalar field $\Phi$ for the double-null
system
\begin{equation}
   \partial_t (\partial_x \Phi +a\Phi)=\partial_y^2 \Phi-2b \partial_y \Phi +S,
     \quad \Phi(0,x,y)=f(x,y),
   \label{eq:simple}
\end{equation}
and the null-timelike system
\begin{equation}
   \partial_t (\partial_x \Phi +a\Phi)=(\partial_x^2 +\partial_y^2)\Phi-2b \partial_y \Phi +S,
    \quad \Phi(0,x,y)=f(x,y),
   \label{eq:simple2}
\end{equation}
with smooth initial data $f(x,y)$ and source $S(t,x,y)$, where $a$ and $b$ are
real constants. Both the $t$ and $x$ directions are characteristic (lghtlike or
null) for (\ref{eq:simple}), whereas the $t$ direction is timelike for
(\ref{eq:simple2}). We include the $b\partial_y \Phi$  term to illustrate how
lower order terms enter the analysis. Note that the systems (\ref{eq:simple})
and (\ref{eq:simple2}) are not transformable into each other and have
essentially different stability properties (compare (\ref{eq:rsteo}) and
(\ref{eq:dampr}) below).

In~\cite{wpsw} the well-posedness of these problems, for $a>0$, was established
by two methods: Fourier-Laplace theory and the energy method. Here we apply
these methods to establish the stability of the discretized version of these
problems. We approximate (\ref{eq:simple}) on a uniform spatial grid
$(x_{j_1},y_{j_2})$  by
\begin{equation}
  {{\partial}\over{\partial t}}(D_{0x}\Phi + a \Phi) = D_{+y}D_{-y} \Phi - 2 b D_{0y} \Phi ,
\label{eq:Dsimple}
\end{equation}
and  approximate (\ref{eq:simple2}) by
\begin{equation}
  {{\partial}\over{\partial t}}(D_{0x}\Phi + a \Phi) = ( D_{+x}D_{-x} 
       +D_{+y}D_{-y}) \Phi - 2 b D_{0y} \Phi ,
\label{eq:Dsimple2}
\end{equation}
in terms of the second order accurate centered ($D_0$) and sidewise  ($D_{\pm}$)
finite difference operators~\cite{kgust} acting on the grid function
$\Phi(t,j_1,j_2)=\Phi(t,x_{j_1},y_{j_2})$. For this Cauchy problem, we keep the
time variable continuous and use the method of lines to treat the resulting
system of ordinary differential equations.

\subsection{The Fourier-Laplace method}
\label{sec:fourlap}

The analysis given in~\cite{wpsw} showed that both the double-null and
null-timelike problems are well posed for $a>0$. Although there are growing
modes, for any initial data $f$ there is a constant $\delta$ which does not
depend upon $\omega_1$ and $\omega_2$ such that $\Re{s} \le \delta$,  so that
the growth rate is bounded. For $a<0$, both problems are ill-posed.

The treatment of the continuum problem can be based upon the Fourier
decomposition
\begin{equation}
   \Phi(t,x,y) = {{1}\over{2\pi}}\int_{\omega_1
   =-\infty}^{+\infty}\int_{\omega_2=-\infty}^{+\infty}
     \hat \Phi(\omega_1,\omega_2,t)e^{i(\omega_1x
    +\omega_2y)}d\omega_1 d\omega_2)
\label{eq:F2D}
\end{equation}
on the domain $(-\infty<(x,y)<+\infty)$. First consider the double-null problem
(\ref{eq:simple}) with source $S=0$. After substituting (\ref{eq:F2D}) in 
(\ref{eq:simple}), we obtain the Fourier space evolution equation
\begin{equation}
    \partial_t  \hat \Phi = s \hat \Phi
\label{eq:FTsimple}
\end{equation}
where 
\begin{equation}
     s=-{{\omega_2^2 +2ib \omega_2 }\over{a+i\omega_1 }}
\label{eq:steo}
\end{equation}
and 
\begin{equation}
   \Re  s=-{a\omega_2^2 +2b \omega_1\omega_2 \over a^2+\omega_1^2 }.
       \label{eq:rsteo}
\end{equation}
Thus for
\begin{equation}
     |2b \omega_1|>  | a\omega_2 |  
       \label{eq:unst0}
\end{equation}
there are growing modes with $\Re s>0$. However, for $a>0$,
\begin{equation}
   \Re  s=-{a\omega_2^2 +2b \omega_1\omega_2 \over a^2+\omega_1^2 }
       =-{a\bigg(\omega_2 +(b/a) \omega_1\bigg)^2 \over a^2+\omega_1^2 }       
         +{b^2\omega_1^2\over a(a^2+\omega_1^2)}
          \le{ b^2 \over a}.
\label{eq:rsteo2}
\end{equation}
so that the exponential growth is bounded independently of $(\omega_1,\omega_2)$
and the problem is well posed.

For the null-timelike problem (\ref{eq:simple2}),
\begin{equation}
   \Re  s=-{a({\omega_2^2 +  \omega_1^2) 
    +2b \omega_1\omega_2}\over{a^2+\omega_1^2 }}.
\label{eq:dampr}
\end{equation}
In this case there are no growing modes for $a>0$, provided $|b|\le a$. For
$a<0$, $ \Re  s \rightarrow +\infty$ as $\omega_2 ^2 \rightarrow +\infty$ with
$\omega_1$ fixed so that there is unbounded exponential growth and the problem
is ill posed.

In order to establish the analogous finite difference results, we consider the
discretized equation (\ref{eq:Dsimple}) on a 2-dimensional periodic grid
\begin{equation}
    x_{j_1} = hj_1,~y_{j_2}=hj_2,~ ~(j_1,j_2)=[0,N-1],~~h=2\pi/N
\label{eq:2Dgrid}
\end{equation}
for the grid function
$$\Phi(t)_{j_1,j_2} =\Phi(t,x_{j_1},y_{j_2})
$$
with $2\pi$-periodicity,
\begin{equation}
\Phi(t,x,y)=\Phi(t,x+2\pi,y+2\pi)=>\Phi(t,{{j_1},j_2})=\Phi(t,{{j_1+N},j_2+N}).
\label{eq:percond}
\end{equation}
After representing $\Phi$ in terms of the discrete Fourier transform
\begin{equation}
      \Phi(t, j_1,j_2) =
      {{1}\over{N}}\sum_{\omega_1=0}^{N-1}\sum_{\omega_2=0}^{N-1}
          \hat \Phi(\omega_1,\omega_2,t)e^{ih (\omega_1j_1+\omega_2j_2)}\
\label{eq:DF2D}
\end{equation}
and substituting in  (\ref{eq:simple}),
we obtain the discrete Fourier space evolution equation
\begin{equation}
   \partial_t \hat \Phi = s \hat \Phi,
\label{eq:Discrete}
\end{equation}
where now
\begin{equation}
          s = -\left( {2 -e^{ih\omega_2} - e^{-ih\omega_2}\over{h}}
    +b(e^{ih\omega_2}  - e^{-ih\omega_2})\right) /({e^{ih\omega_1}
    - e^{-ih\omega_1}\over{2}} + ah).
\label{eq:sexp}
 \end{equation}
 
With the notation
 \begin{equation}
 \omega_{10} = {{\sin(h\omega_1)}\over{h}}, \quad \omega_{20} =
        {{\sin(h\omega_2)}\over{h}},
      \quad \omega_{21} = {{\sin(h\omega_2/2)}\over{h/2}} ,
 \end{equation}
(\ref{eq:sexp}) takes a form similar to the analytic expression~(\ref{eq:steo}),
\begin{equation}
s=-{{\omega_{21}^2 +2ib \omega_{20} }\over{a+ i\omega_{10} }}.
\label{eq:simp}
\end{equation}
Therefore
\begin{eqnarray}
\Re s &=& -{a\omega_{21}^2+2b\omega_{20}\omega_{10}\over a^2+\omega_{10}^2}
 \nonumber \\
&=&-\bigg(4a\sin^2({\omega_2h\over 2})+2b\sin( \omega_2h) \sin(\omega_1h) \bigg ) /
\bigg (h^2a^2+\sin^2(\omega_1h) \bigg ) .
\label{eq:7}
\end{eqnarray}

The stability of the finite difference algorithm requires that there is a
constant $\delta > 0$ such that $\Re s\le \delta$ for all $(\omega_1 h,\omega_2
h)$~\cite{kgust}. Since the numerator of  (\ref{eq:7}) is bounded, an
examination of the denominator of (\ref{eq:7}) shows that this is certainly true
if there is a constant $\delta_1>0 $ such that  $|\sin \omega_1 h|\ge \delta_1$
for all $\omega_1h$. Thus we can reduce the analysis to the case that there is a
sequence  $\omega_1 h$ such that $\sin \omega_1 h \to 0.$ Then an examination of
the numerator of (\ref{eq:7}) shows  $\Re s<0$ if $ |{\omega_2 h\over 2}|\ge
\delta_2 >0.$ Thus we can further reduce the analysis to the case that there is
a sequence such that  $\omega_2 h \to 0$. In that limit we can simplify
(\ref{eq:7}) to
\begin{eqnarray}
\Re s&=&-\bigg (a\omega_2^2+2b\omega_2{\sin( \omega_1 h)\over h} \bigg ) /
      \bigg ( a^2+{\sin^2 (\omega_1 h)\over h^2} \bigg )
\nonumber \\
&=& \bigg (-a[\omega_2+{b\over a}{\sin (\omega_1h)\over h}]^2 
+{{b^2\over a}{\sin^2( \omega_1h)\over  h^2} \bigg ) /
\bigg ( a^2+{\sin^2\omega_1 h\over h^2}}\bigg ).
\end{eqnarray}
Now, for $a>0$, 
\begin{equation}
   \Re s \le {b^2\over a}.
\end{equation}
This is the discrete version of the continuum result in~\cite{wpsw} and
establishes stability of the finite difference algorithm for $a>0$. For $a< 0$
the continuum problem is ill posed.

\bigskip

\subsection{The energy method}
\label{sec:energy} 

For the generalization of numerical stability to variable coefficients and
quasilinear problems, it is necessary to show that the finite difference
equation (\ref{eq:Dsimple}) is stable against lower order perturbations. For
that purpose, we first establish an energy estimate.

We denote by $(\Phi,\Psi)_h $ and $ \|\Phi\|_h $ the usual discrete version of
the $L_2$ scalar product and norm. Also, we employ the standard $D_{0}$ and
$D_{\pm}$ finite difference operators which obey the summation by parts rules so
that periodicity guarantees that there are no boundary terms. We denote
$\partial_t \Phi =\Phi_t$.

We first consider the semi-discrete equation \begin{equation}
\partial_t(D_{0x}\Phi+a\Phi)=D_{+y}D_{-y} \Phi-2bD_{0y} \Phi+S, \quad a>0.
\label{eq:dssimple} \end{equation} We have
$$
      \left(\Phi,\partial_t(D_{ox}\Phi+a\Phi)\right)_h
    =-\|D_{+y}\Phi\|^2_h +(\Phi,S)_h
$$
or
$$
 (\Phi,D_{0x}\Phi_t)_h+{a\over 2}\partial_t \|\Phi\|^2_h=-\|D_{+y}\Phi\|^2_h +(\Phi,S)_h ,
$$
which gives
\begin{equation}
   {a\over2}\partial_t\|\Phi\|^2_h+\|D_{+y}\Phi\|_h^2
      =(D_{0x}\Phi,\Phi_t)_h +(\Phi,S)_h.
     \label{eq:9}
\end{equation} 

Next,    
$$  
    \left(\Phi_t,\partial_t (D_{0x}\Phi+a\Phi)\right)_h
    =(\Phi_t,D_{+y}D_{-y}\Phi)_h-2b(\Phi_t,D_{0y}\Phi)_h
    + (\Phi_t,S)_h,
$$
which gives
\begin{equation}   
    a\|\Phi_t\|^2_h+\frac{1}{2} \partial_t \|D_{+y}\Phi\|^2_h=-2b(\Phi_t,D_{0y}\Phi)_h + (\Phi_t,S)_h .
    \label{eq:10}
\end{equation}    
  
Next,
$$
\left(D_{0x}\Phi,\partial_t (D_{0x}\Phi+a\Phi)\right)_h
     =(D_{0x}\Phi,D_{+y}D_{-y}\Phi)-2b(D_{0x}\Phi,D_{0y}\Phi)
      + (D_{0x}\Phi,S)_h ,
$$
which gives
\begin{equation}
   \frac{1}{2}\partial_t\|D_{0x}\Phi\|^2_h
        =-2b(D_{0x}\Phi,D_{0y}\Phi)-(D_{0x}\Phi,a\Phi_t)_h  
      + (D_{0x}\Phi,S)_h.
   \label{eq:11}
\end{equation}

Addition of (\ref{eq:9})--(\ref{eq:11}) gives
\begin{eqnarray}
        \partial_t \bigg ({a\over 2}\|\Phi\|^2_h &+&\frac{1}{2} \|D_{+y}\Phi\|^2_h
     +\frac{1}{2} \|D_{0x}\Phi\|^2_h \bigg  )
    +\|D_{+y}\Phi\|^2+a\|\Phi_t\|_h^2  \nonumber \\
        &=&(1-a)(D_{0x}\Phi,\Phi_t)_h-2b(\Phi_t,D_{0y}\Phi)_h
        -2b(D_{0x}\Phi,D_{0y}\Phi)_h
        + (\Phi+\Phi_t+D_{0x}\Phi,S)_h.
\end{eqnarray}
Now standard inequalities, e.g. $2(\Phi,\Psi)_h\le  \|C\Phi\|^2_h+
\|C^{-1}\Psi\|^2_h$, lead to the finite difference version of the energy
estimate for the continuum problem obtained in~\cite{wpsw},
\begin{eqnarray}
\frac{1}{2}\partial_t \bigg (a\|\Phi\|^2_h &+& \|D_{+y}\Phi\|^2_h+ \|D_{0x}\Phi\|^2_h \bigg  )
    +\|D_{+y}\Phi\|^2_h+\frac{3a}{8}\|\Phi_t\|_h^2  \nonumber \\
        &\le& const\bigg (\|\Phi\|^2_h + \|D_{+y}\Phi\|^2_h
        +\|D_{0x}\Phi\|^2_h + \|S\|^2_h \bigg  ) ,
\end{eqnarray}
which controls the growth of $\Phi$ and its derivatives. 

As in the continuum problem, the same technique shows that the finite difference
equation
\begin{equation}
  \partial_t(D_{0x}\Phi+a\Phi)=(D_{+x}D_{-x}+ D_{+y}D_{-y} )\Phi-2bD_{0y}\Phi 
     +cD_{0x}\Phi   +d\partial_t  \Phi   + e \Phi +S
   \label{eq:lowsw}
\end{equation}
has an energy estimate provided the lower order terms satisfy $(a-d)>0$. In
addition, estimates for the higher derivatives of $\Phi$ follow from the
equations obtained by differentiating  or finite differencing (\ref{eq:lowsw}).
These properties are sufficient to guarantee the stability of the corresponding
evolution problem with variable coefficients.

\section{Simulation of the Cauchy problem}
\label{sec:testper} 

In order to test and explore the properties of the characteristic evolution
algorithms described in Sec.~\ref{sec:alg}, we implement a code based upon a
2-dimensional discrete Fourier transform on a periodic $(x,y)$ grid, with a
$4^{th}$ order Runge-Kutta time integrator. For convenience, we carry out the
simulations  in the domain $(x,y)\in[0,1)$, with the appropriate rescaling of
the conventions used in  Sec.~\ref{sec:alg}. Except for convergence tests, the
simulations were carried out on a basic  grid of size $N\times N=64^2$, with
$h=\Delta x =1/N$ and timestep $\Delta t = h/5$, corresponding to a Courant
factor $\Delta t/ \Delta x =1/5$. 

Periodicity in the $x$-direction implies that the $x$-axis is a closed lightlike
line. The simulations are performed in order to test and analyze the resulting
behavior of the wave equations (\ref{eq:simple}) and (\ref{eq:simple2}), and are
divided between (i) source free evolution with non-zero initial data and (ii)
evolution with non-vanishing source and vanishing initial data.

For the first set of simulations, we prescribe initial data consisting of a
compact pulse 
\begin{equation}
    \Phi(0,x,y) = f(x,y) = A x^4y^4(x-1)^4(y-1)^4,
\label{eq:compactF}
\end{equation}
and vanishing source $S=0$, with amplitude $A=10^4$, which
normalizes the magnitude close to unity.

For the second set of simulations, we prescribe vanishing initial data
$ \Phi(0,x,y)=0$  and switch a compact source on and off according to
\begin{eqnarray}
\label{eq:compactS}
S &=& A\partial_t \left [ (t-t_1)(x-x_1)(y-y_1) (t-t_2)(x-x_2)(y-y_2)\right ]^4 , 
              \quad t_1\le t \le t_2,  \, x_1\le x \le x_2,  \, y_1\le y \le y_2 ,\\
\nonumber
S &=& 0, \quad  {\rm otherwise},         
\end{eqnarray}
where the amplitude is again $A=10^4$, for the same consideration. The source is
turned on at $t_1=2$, and shut off at $t_2=8$, with peak amplitude at $t=4$. The
spatial dependence of the pulse is compactified between $x_1=y_1=0.2$, and
$x_2=y_2=0.8$.

\subsection{Convergence Tests}

\begin{figure}[ht]
\centering
\includegraphics*[width=10cm]{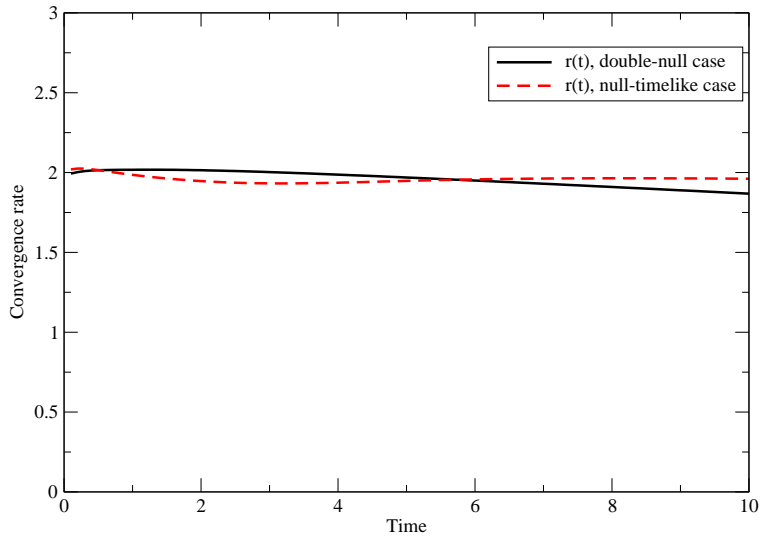}
\caption{Convergence rate $r(t)$, in the interval $0\le t \le 10$, of the
numerical solution $\Phi$ for the double-null case (\ref{eq:simple}) and
null-timelike case (\ref{eq:simple2}), with $b=0.1$ and $a=1$.}
\label{fig:ConvPlot}
\end{figure}

Convergence tests were based upon three grids of size $N_{A}\times N_{A}$, with 
$N_A=64/A$, in the ratio  $A=(1,2,4)$,  $h_A=Ah$, while keeping the Courant
factor fixed at $\Delta t_A / h_A =1/5$. The Cauchy convergence rate $r$ was
found in terms of  the corresponding computed solutions $\Phi_A(t,j_1,j_2)$
according to the formula
\begin{equation}
    r(t)=\frac{\log(||\Phi_{4}-\Phi_{2})||_h/||\Phi_{2}-\Phi_{1}||)_h}{\log(2)}.
\label{eq:Convfact}
\end{equation}
In Fig.~\ref{fig:ConvPlot} we plot the measured convergence rate
(\ref{eq:Convfact}) for the three grids in the interval $0\le t \le 10$.  The
convergence rate $r(t)\approx2$ is in excellent accord with that expected from
the second order finite difference approximations (\ref{eq:Dsimple})
and~(\ref{eq:Dsimple2}). At late times, the convergence rate for the double-null
problem slowly degrades because of the inability of the coarsest $16\times 16$ 
grid to resolve the high frequency growing modes (\ref{eq:unst0}).

\subsection{Stability Tests}

Test runs for the two systems (\ref{eq:simple}) and~(\ref{eq:simple2}) confirmed
the results for the well-posedness of the analytic problem found in~\cite{wpsw}:

\begin{itemize}

\item {For $a < 0$, both problems are ill posed. Numerical instability  is
evident and the runs quickly crash}.

\item For the double-null system (\ref{eq:simple}) with $a=1$ and $|b|>0$, there
are exponentially growing modes but the runs are numerically stable and
convergent.

\item For the null-timelike system (\ref{eq:simple2}) with $a=1$ and $|b| < a$,
there are no growing modes. For $|b|>a$, there are exponentially growing modes,
but the runs are numerically stable and convergent.

\item In all simulations for $a>0$, the wave remains smooth and there is no sign
of numerical instability.

\end{itemize}

\begin{figure}[ht]
\centering
\includegraphics*[width=8cm]{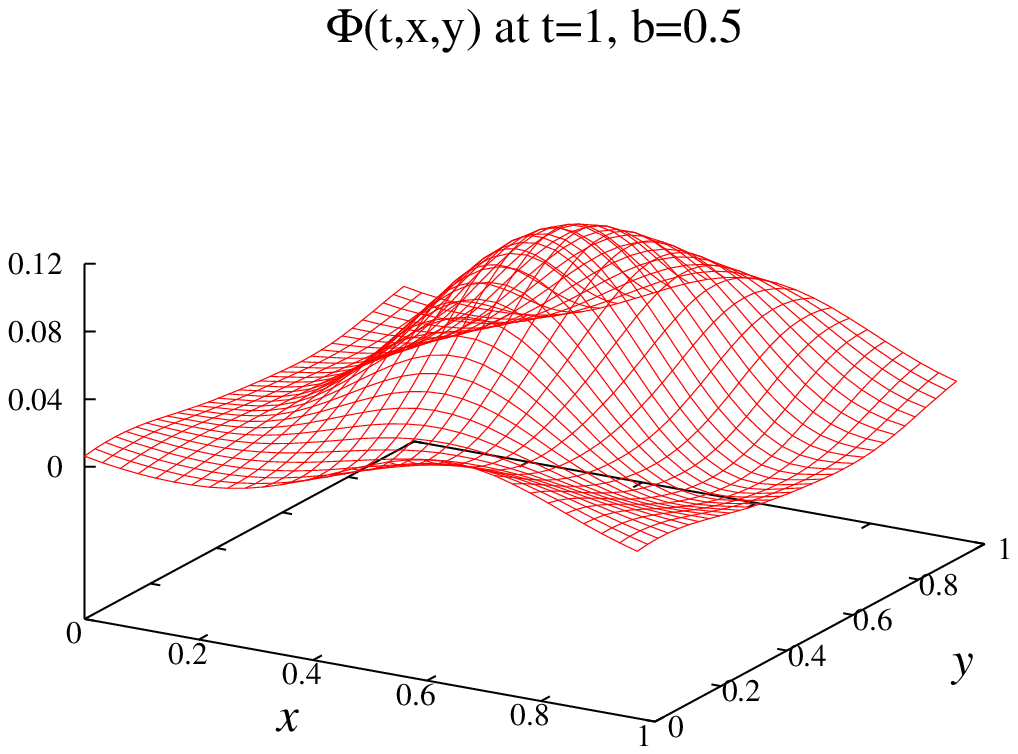}
\includegraphics*[width=8cm]{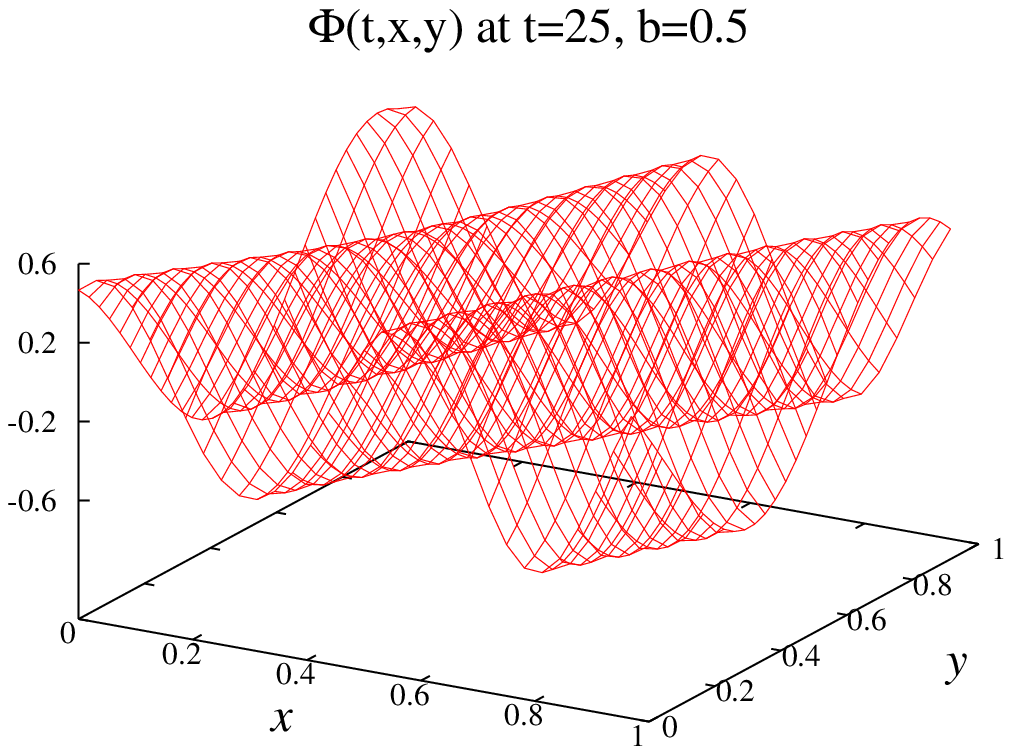}
\caption{Snapshots of the simulation of the double-null system (\ref{eq:simple}),
with initial pulse (\ref{eq:compactF}),  for $b=0.5$ and $a=1$. On the left, at
$t=1$, the growing mode is barely visible. On the right, at $t=25$, the growing
mode dominates but the smoothness indicates no sign of a numerically induced
instability. The case $b=1$ shows more extreme growth.}
\label{fig:Eq1b05}
\end{figure}

Figure \ref{fig:Eq1b05} displays the effect of the lower
order terms for the evolution of the double-null system  with $a=2b=1$,
for the initial pulse (\ref{eq:compactF}) and source $S=0$. The fastest
growing mode in the analytic problem is not evident at the early time $t=1$ but
becomes dominant at $t=25$. The solution continues to grow
but remains smooth, showing no sign of a numerically induced instability.
There is more extreme growth for the case $a=b=1$. 

\begin{figure}[ht]
\centering
\includegraphics*[width=8cm]{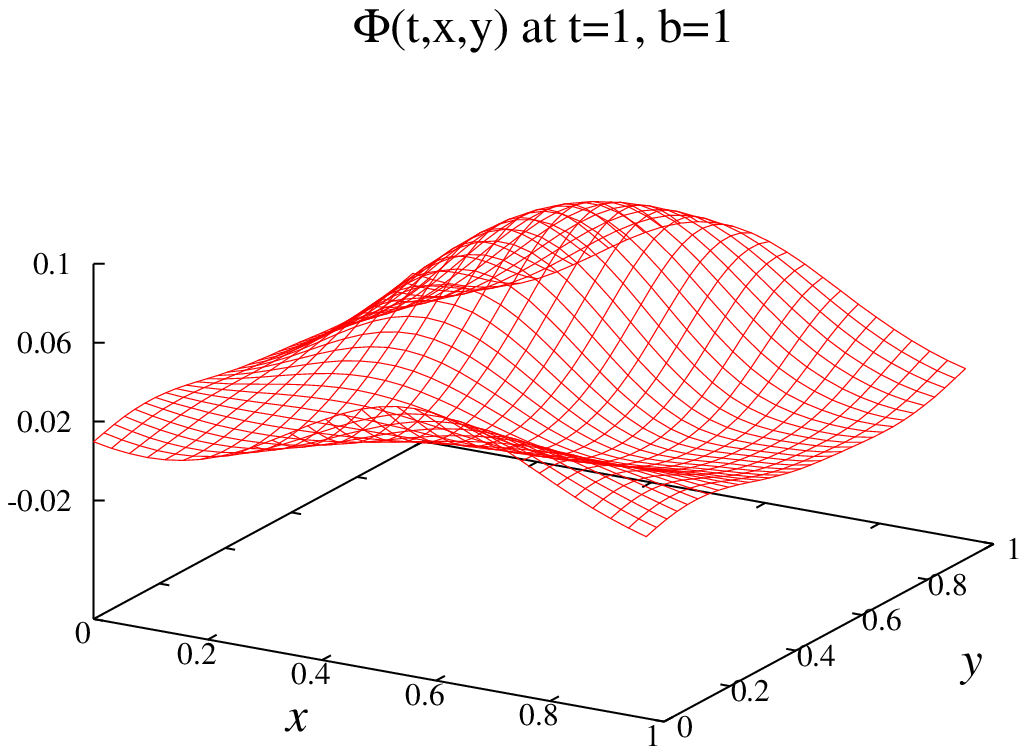}
\includegraphics*[width=8cm]{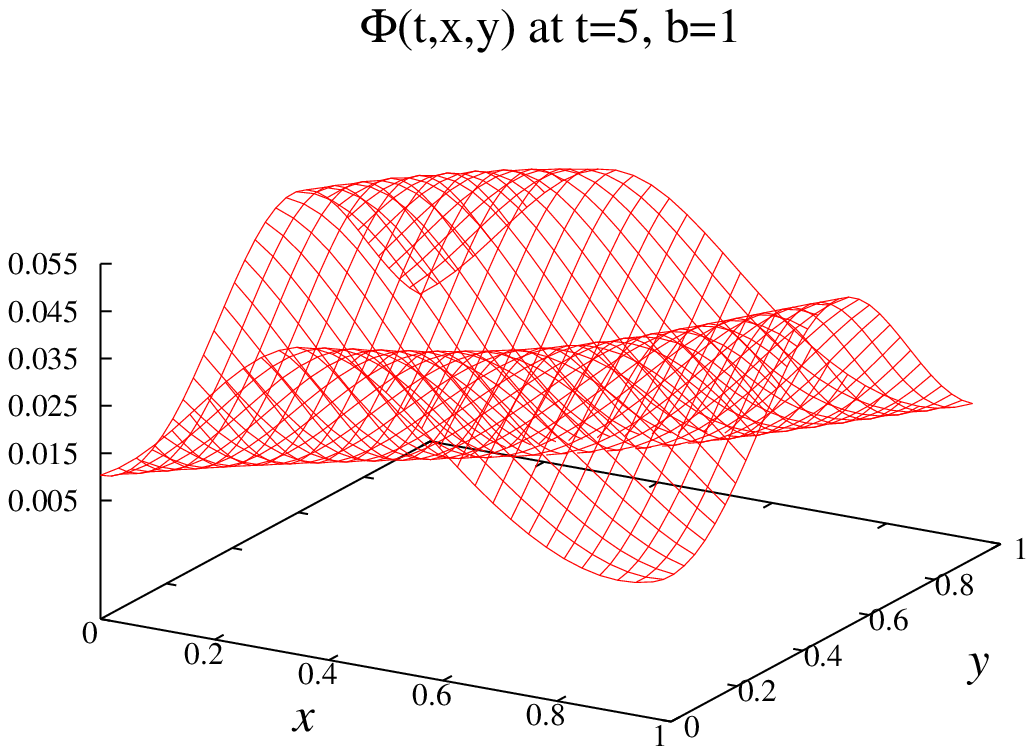}
\caption{Snapshots of the simulation of the null-timelike system (\ref{eq:simple2}),
with initial pulse (\ref{eq:compactF}) for the neutrally stable case $b=a=1$ At
$t=1$ (left), the pulse has begun to develop asymmetry. At $t=5$ (right), the
critical mode with symmetry along the diagonal lines has begun to form.}
\label{fig:Eq2b1}
\end{figure}

\begin{figure}[ht]
\centering
\includegraphics*[width=8cm]{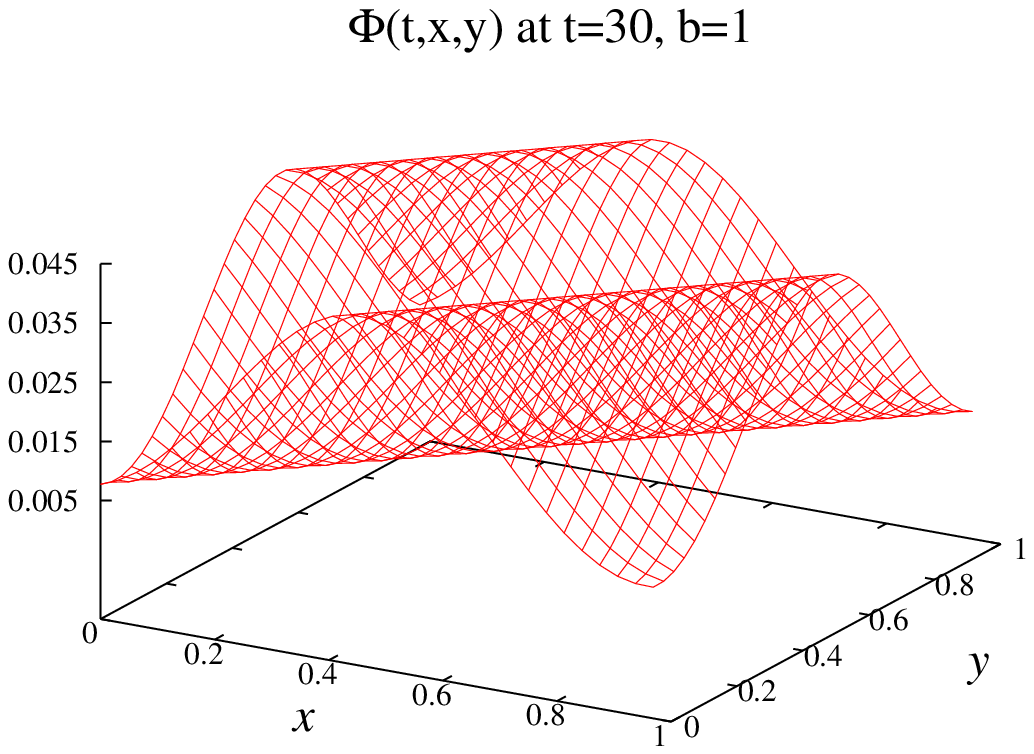}
\includegraphics*[width=8cm]{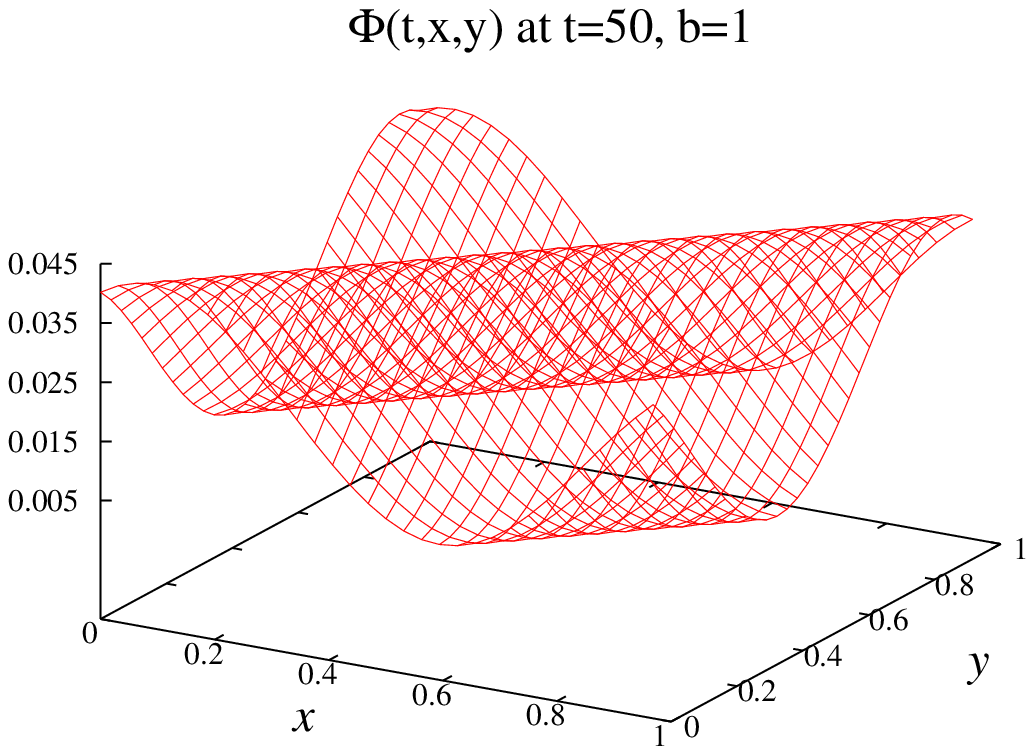}
\caption{Late time behavior of the null-timelike system (\ref{eq:simple2}), with
initial initial pulse (\ref{eq:compactF}), for the neutrally stable case
$b=a=1$. The late time behavior at $t=30$ (left) and $t=50$ (right)  shows that
critical mode maintains a constant amplitude as the wave travels.}
\label{fig:Eq2b50}
\end{figure}

For the null-timelike system (\ref{eq:simple}), the quasi-normal modes  
\begin{equation}
     \Phi = e^{st +i(\omega_1x+\omega_2 y)}
     \label{eq:mode}
\end{equation}
have growth (decay) rate $\Re s$ given by (\ref{eq:dampr}). Consequently, $\Re
s<0$ for $|b|<a$ and the modes are damped. Simulations for $a=1$ and $b=0.5$
show no sign of growth and the wave decays at late times. For $b>a$, 
(\ref{eq:dampr}) implies that the dominant growing mode satisfies $\omega_2 =-
b\omega_1/a$ so that the wave develops a symmetry along the lines $x-(b/a)y
=const$. For the neutrally stable case  $a=b=1$ illustrated in
Fig.~\ref{fig:Eq2b1},  initial growth of the critical mode with $\omega_2 =-
\omega_1$ is evident at $t=5$.  As shown in Fig.~\ref{fig:Eq2b50}, this
neutrally stable mode does not exhibit exponential growth or decay but settles
into a constant amplitude traveling wave. The late time
behavior shows the expected symmetry along the diagonal lines $x-y=const$. (For
$b=-a$ the symmetry switches to the diagonals $x+y=const$.)

\clearpage

\subsection{Simulations of the double-null problem}

\begin{figure}[ht]
\centering
\includegraphics*[width=8cm]{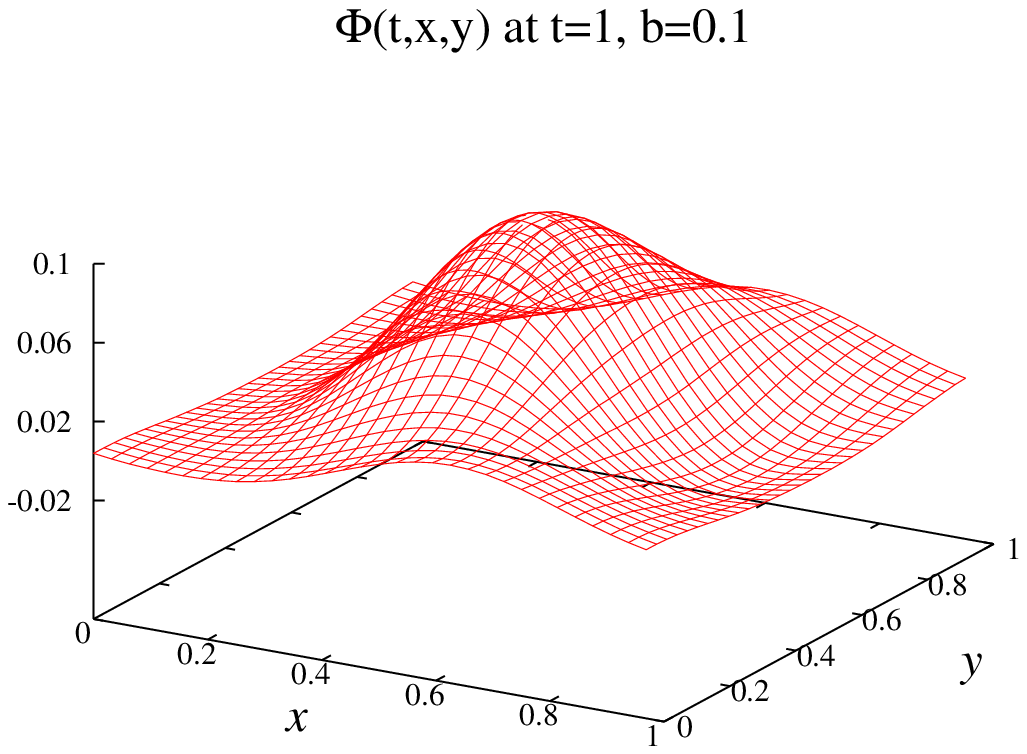}
\includegraphics*[width=8cm]{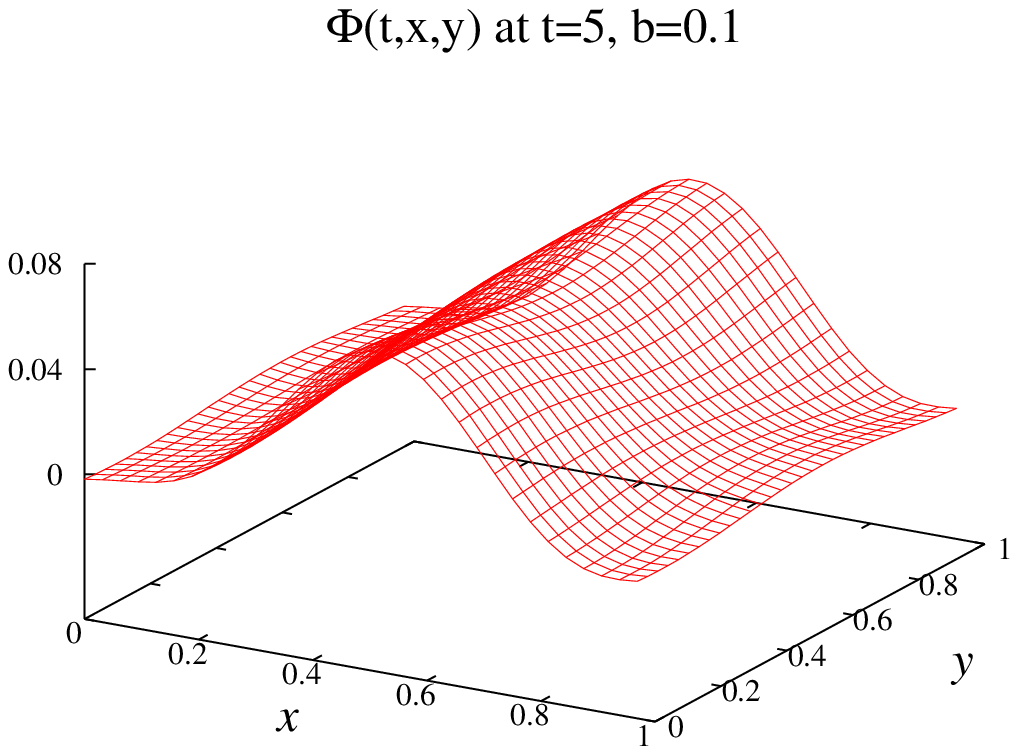}
\caption{Early time simulation of the double-null problem for $b=0.1$ and $a=1$
with initial data (\ref{eq:compactF}) and source $S=0$.  At $t=1$ (left plot)
the wave has undergone little change but at $t=15$ (right plot) it begins to
homogenize in the $y$=direction.}
\label{fig:reFcompact1}
\end{figure}

\begin{figure}[ht]
\centering
\includegraphics*[width=8cm]{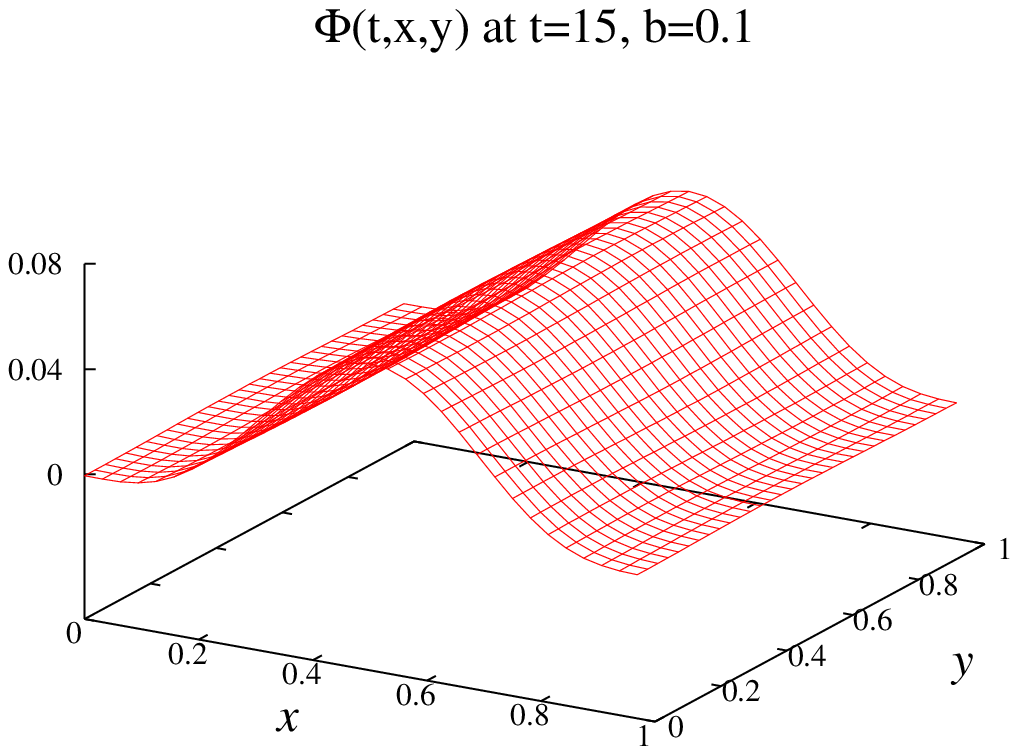}
\includegraphics*[width=8cm]{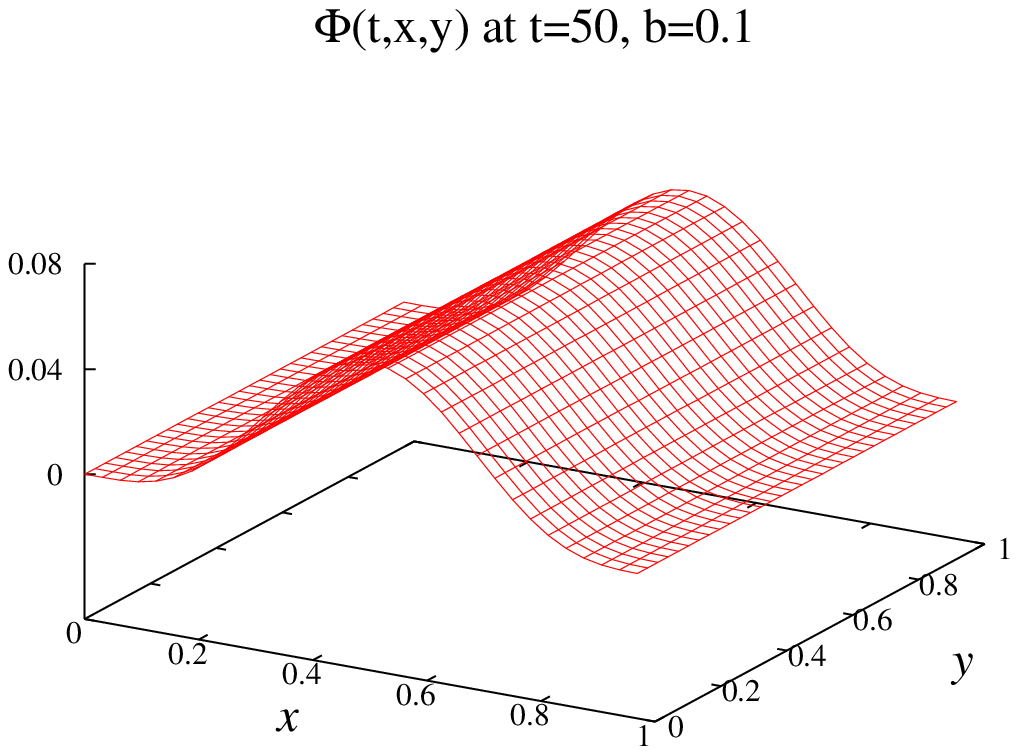}
\caption{Late time simulation of the double-null problem for $b=0.1$ and $a=1$
with initial data (\ref{eq:compactF}) and source $S=0$. At $t=15$ (left plot)
the wave has become uniform in the $y$-direction. At this time the wave has 
settled into a stable mode and its shape has frozen, as evident from comparison
with the right plot at  $t=50$.}
\label{fig:reFcompact15}
\end{figure}

Here we present simulations of the double-null problem (\ref{eq:simple}) for
the case $a=1$ and $b =0.1$ where no growing modes are excited. First, for the
compact initial pulse (\ref{eq:compactF}) with source $S=0$,
Fig's.~\ref{fig:reFcompact1} and \ref{fig:reFcompact15} show that two
qualitative features emerge on the order of a few crossing times: the
$y$-dependence become uniform and the wave freezes in shape. This behavior can
be inferred from the integral
\begin{equation}
\label{eq:eestim}
    \int dx dy  \Phi_t \bigg ( \partial_t (\Phi_x+a\Phi) -\partial_y^2 \Phi
         +2b\partial_y \Phi \bigg) =0
         \end{equation}
which implies 
\begin{equation}
\label{eq:eestim2}
    \partial_t\int dx dy( \partial_y \Phi)^2 =-2  a\int dx dy{\Phi_t}^2 
    -4b \int dx dy{\Phi_t}\partial_y \Phi.
    \label{eq:hom}
\end{equation}
Thus, for $b<< a$, the $y$-dependence monotonically becomes homogeneous as
$\Phi_t \rightarrow 0$ and the waveform freezes as it propagates with zero
coordinate velocity along the characteristic in the $t$-direction. 

\clearpage

Next we present simulations with vanishing initial data and a non-vanishing
source. As an indication of what to expect, consider the example
$$S=\delta(x-\frac{1}{2})\partial_t \psi(t),
$$
where $\psi(t)$  is a compact signal in time. The corresponding solution of
(\ref{eq:simple}) is
$$\Phi(t,x) =\frac {e^{-a/2} }{1-e^{-a}} e^{-ax} \psi(t) , \quad 0\le x<\frac{1}{2}$$
$$\Phi(t,x) =\frac {e^{a/2} }{1-e^{-a}}  e^{-ax}\psi(t)  , \quad \frac{1}{2} <x <1 .$$
Notably, a signal emitted in the forward $x$-direction instantaneously circuits
the closed lightlike line and returns to the source. 

\begin{figure}[ht]
\centering
\includegraphics*[width=8cm]{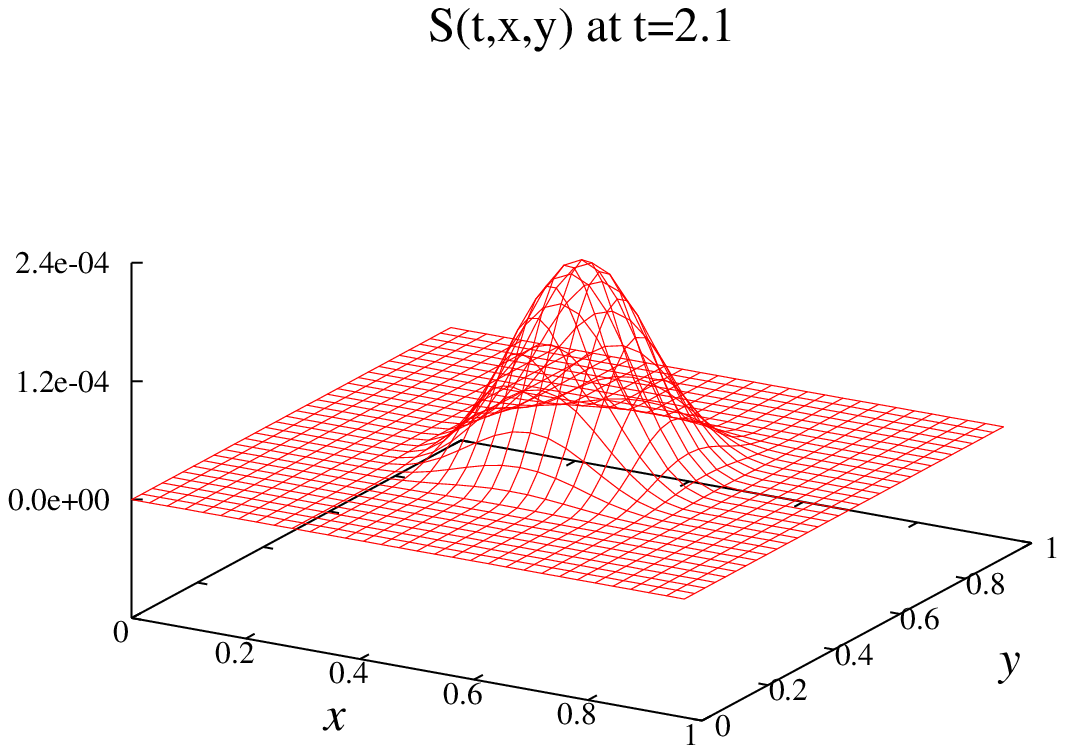}
\includegraphics*[width=8cm]{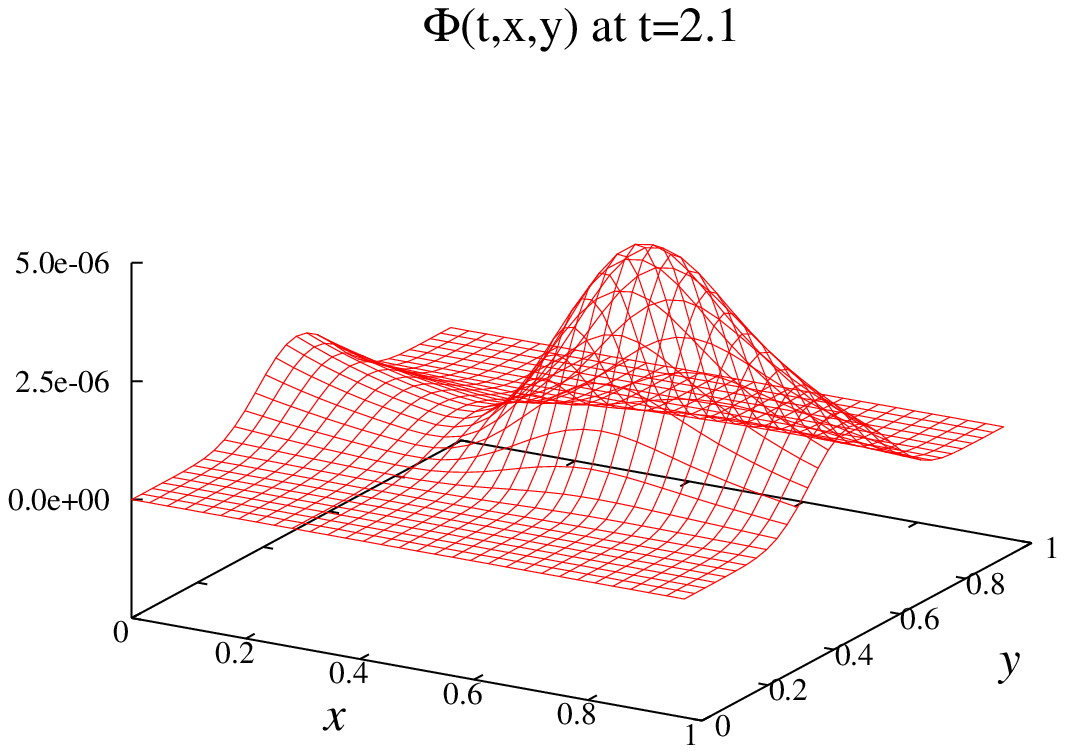}
\caption{Simulation of the double-null problem for $b=0.1$ and $a=0.5$ with
compact source (\ref{eq:compactS}). The left plot shows the shape of the source,
at $t=2.1$, just after it has been turned on. At this same time, the right plot
shows that the resulting wave has instantaneously propagated around the
closed lightlike line in the $x$-direction.}
\label{fig:reFScompact2}
\end{figure}

\begin{figure}[ht]
\centering
\includegraphics*[width=8cm]{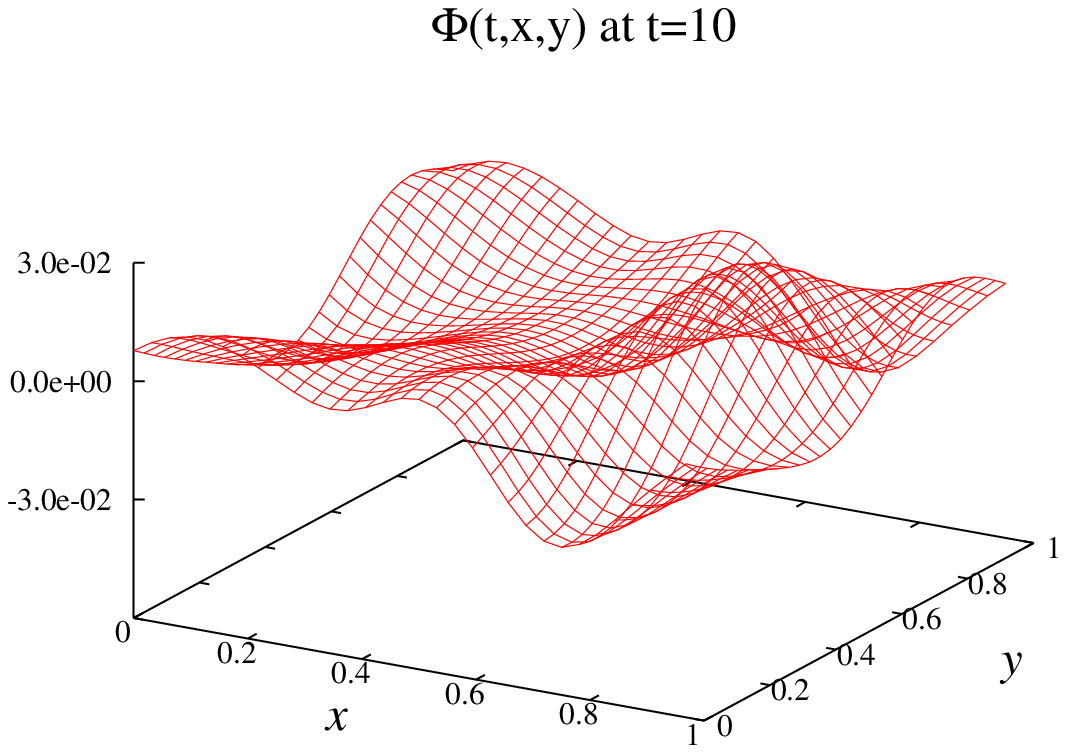}
\includegraphics*[width=8cm]{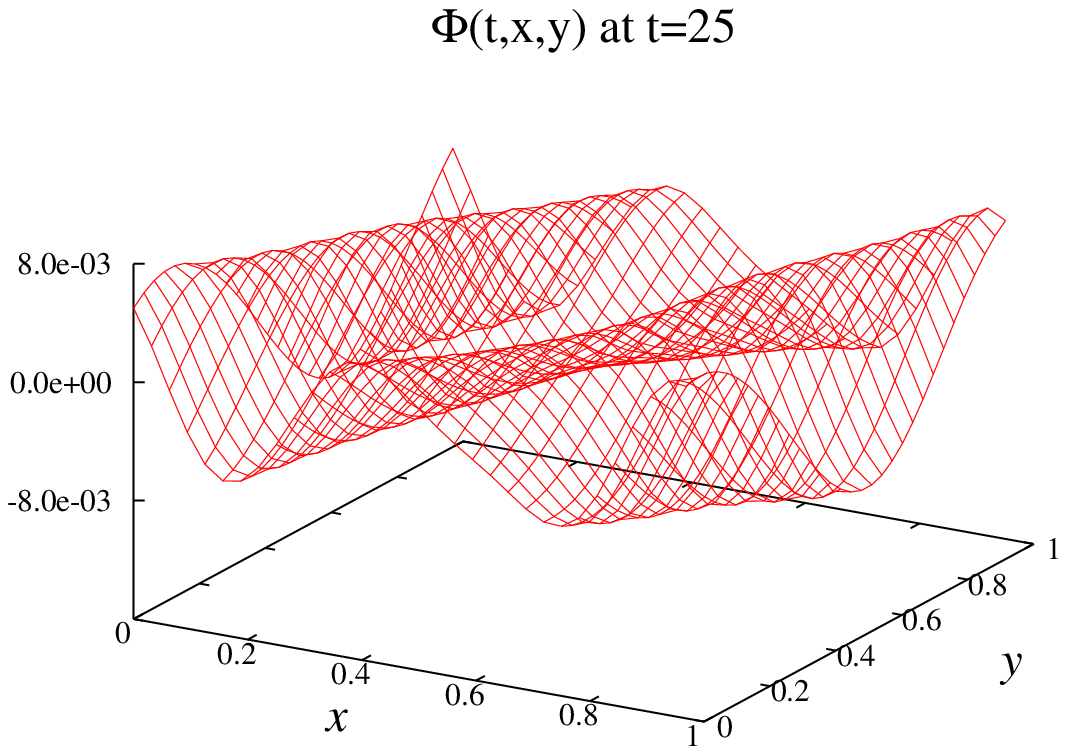}
\caption{The propagation of the signal several crossing times after the source
(\ref{eq:compactS}) has turned off for the double-null problem with $b=0.1$ and
$a=0.5$. At $t=10$ (left plot),  the wave has spread in the $y$-direction and
there is no remaining evidence of the  original compactness of the source.  The
plot at $t=25$ (right) shows that the solution continues to slowly decay as it
becomes more uniform.} \label{fig:reFScompact25}
\end{figure}

Figures \ref{fig:reFScompact2} and \ref{fig:reFScompact25} present snapshots of
the numerical evolution of the double-null problem  with the compact
pulse-shaped source (\ref{eq:compactS}) for $b=0.1$ and $a =0.5$.   In
Fig.~\ref{fig:reFScompact2}, at $t=2.1$ just after the source has turned on, it
is clear that the signal has instantaneously propagated around the closed
lightlike curve in the $x$-direction. The effect of the $a$-term is to damp the
signal as it returns to the source.

After the source is turned off at $t=8$, the wave continues to spread in the
$y$-direction as its amplitude decays. At $t=10$, there is little evidence of
the original compactness of the source. At late times, as shown in
Fig.~\ref{fig:reFScompact25}, the wave continues to decay as it becomes more
uniform. 

\clearpage

\subsection{Simulations of the null-timelike problem}

Here we present simulations of the null-timelike problem (\ref{eq:simple2}), for
the case $a=1$ and $b =0.1$ where no growing modes are present. Now the
$t$-direction is no longer characteristic and there are waves propagating in
both the $\pm x$-directions. For simulations with source $S=0$, an analysis
analogous to that leading to (\ref{eq:hom}) for the double-null problem  gives
\begin{equation}
\label{eq:eestim22}
    \partial_t\int dx dy \bigg( (\partial_x \Phi)^2+ (\partial_y  \Phi)^2\bigg)
     =-2  a\int dx dy{ \Phi_t}^2    -4b \int dx dy{ \Phi_t}\partial_y  \Phi.
       \label{eq:hom2}
\end{equation}

\begin{figure}[ht]
\centering
\includegraphics*[width=8cm]{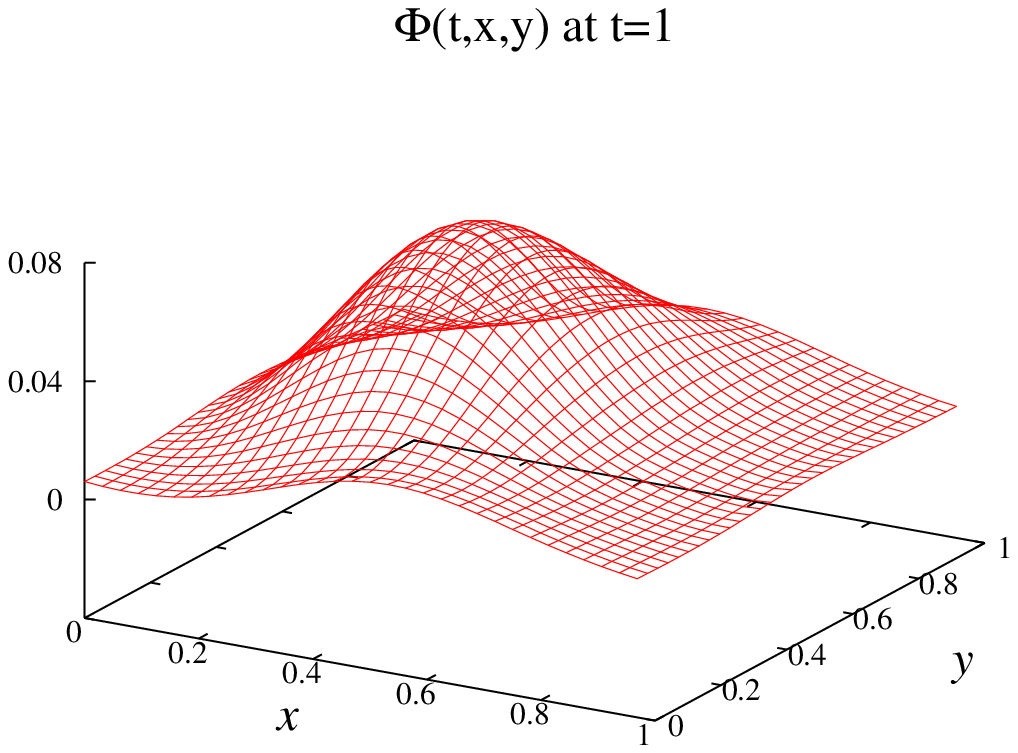}
\includegraphics*[width=8cm]{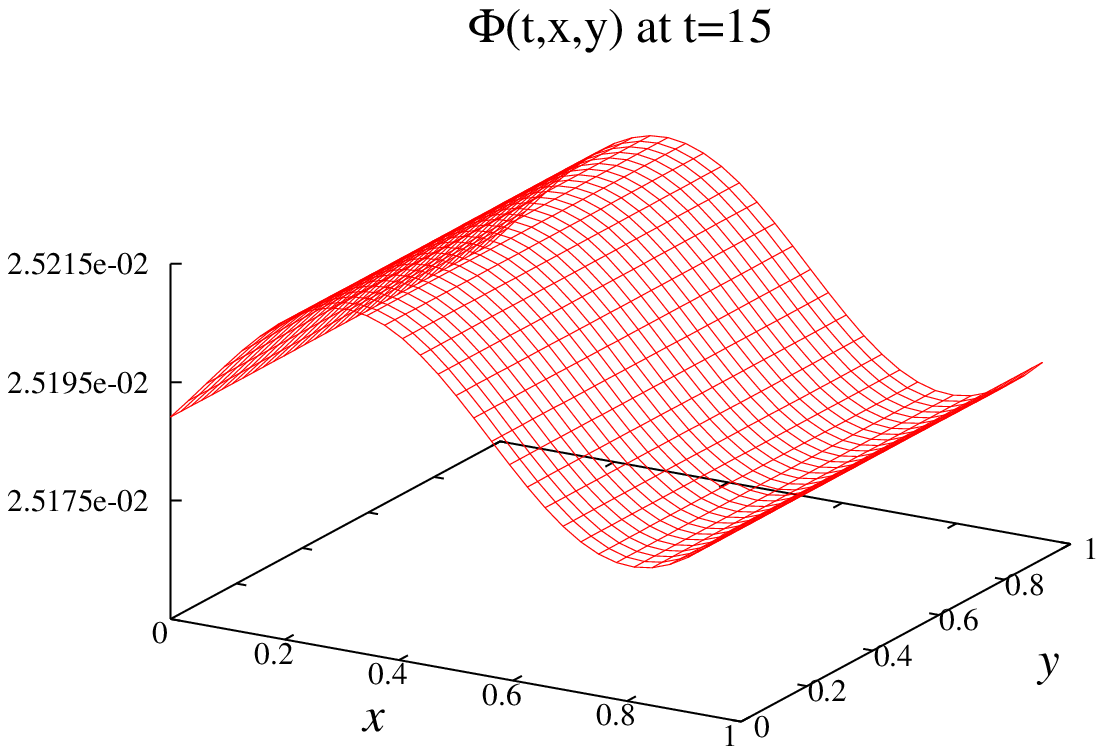}
\caption{Evolution of the null-timelike problem for $b=0.1$ and $a=1$
with initial pulse (\ref{eq:compactF}) and source $S=0$. 
At $t=1$ (left plot) the wave has undergone little change. 
At $t=15$ (right plot) the wave has become uniform in the $y$-direction and
has settled into a long wavelength mode traveling in the $x$-direction.}
\label{fig:reTcompact}
\end{figure}

For $b<< a$, (\ref{eq:hom2}) implies that the amplitude decays monotonically as
$\Phi_t \rightarrow 0$.  As in the double-null problem, the simulations with initial pulse
(\ref{eq:compactF})  show that the $y$-dependence becomes uniform after a few
crossing times, as illustrated in Fig.~\ref{fig:reTcompact}. However, the
$x$-dependence persists as the wave decays, as can
be understood in terms of the damping rate (\ref{eq:dampr}) for the quasi-normal
modes.  After the $y$-dependence has damped out, i.e. as $\omega_2 \rightarrow
0$, (\ref{eq:dampr}) reduces to
\begin{equation}
       \Re s \rightarrow -\frac{a}  {1+(a^2/\omega_1^2)}
\end{equation}
so that the long wavelength $x$-dependence damps very slowly as
$\Phi \rightarrow const$.

\begin{figure}[ht]
\centering
\includegraphics*[width=8cm]{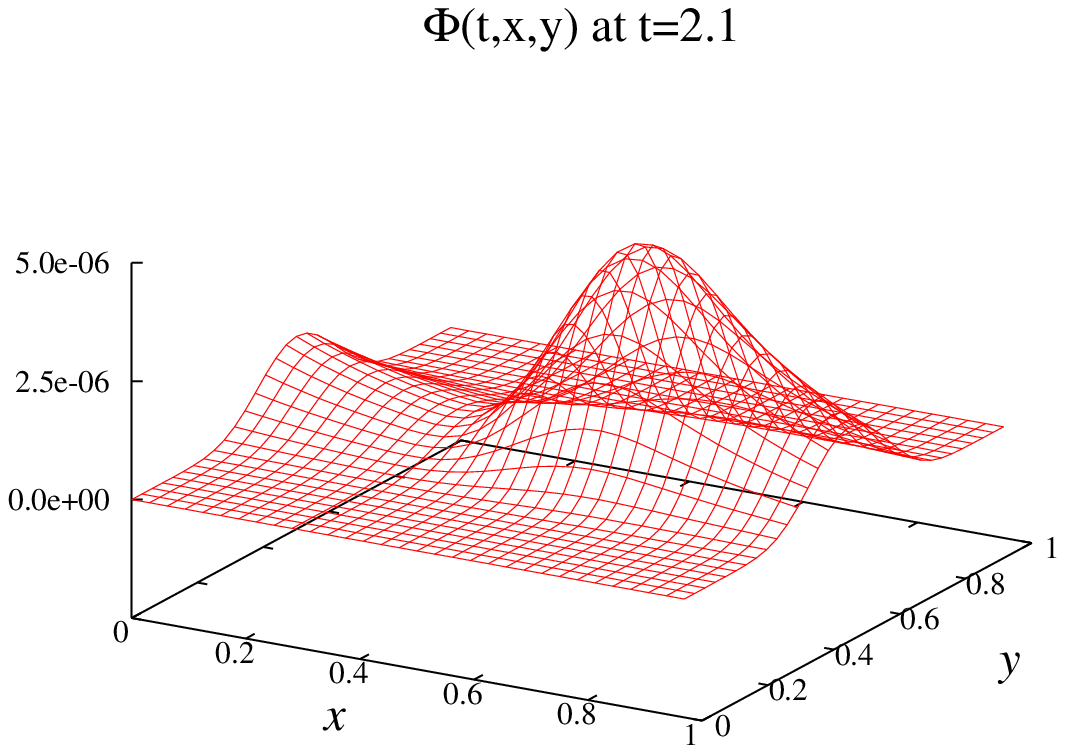}
\includegraphics*[width=8cm]{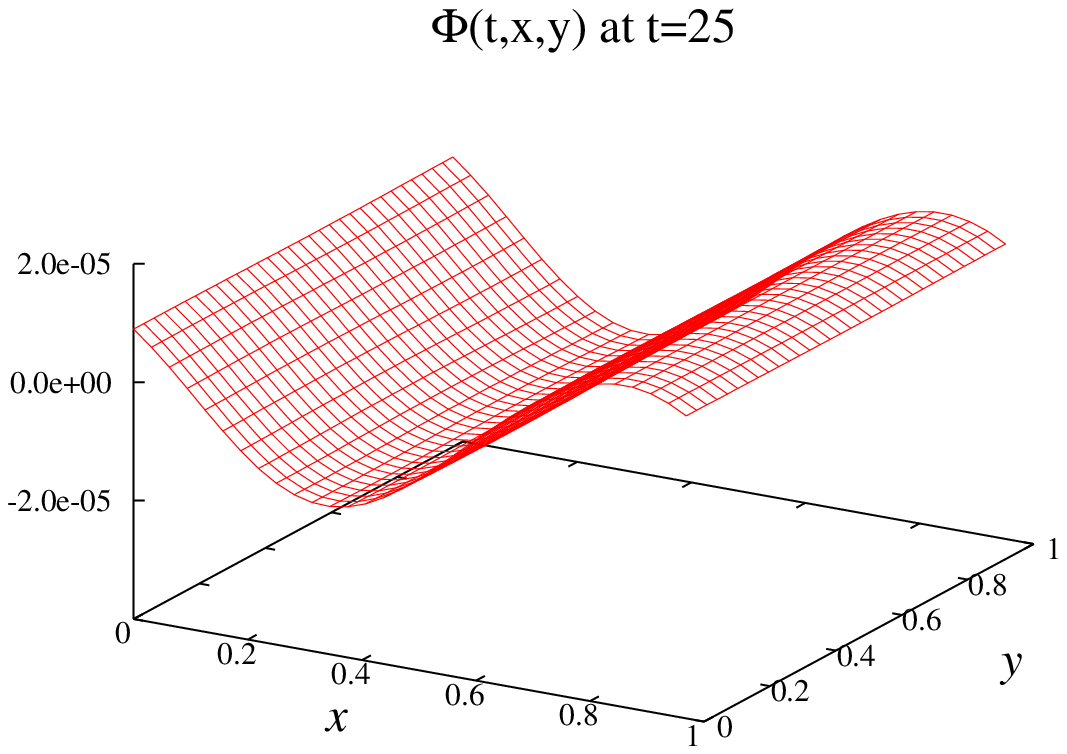}
\caption{Simulation of the null-timelike problem for $b=0.1$ and $a=0.5$ with
compact source (\ref{eq:compactS}).  The left snapshot at $t=2.1$, just after
the source is turned on, shows that the signal has instantaneously propagated
around the closed lightlike line in the $x$-direction, very similar to the
corresponding double-null problem illustrated in Fig.~\ref{fig:reFScompact2}.
The late time behavior in the right snapshot at $t=25$ shows that the wave
homogenizes in the $y$-direction as it decays, but in a quite different manner
than the decay in the double-null problem illustrated in
Fig.~\ref{fig:reFScompact25}.}
\label{fig:reTScompact}
\end{figure}

Next, we present snapshots of the simulation of the null-timelike problem  for
$b=0.1$ and $a = 0.5$ with vanishing initial data and the compact pulse-shaped
source $S$ given by (\ref{eq:compactS}).  The left snapshot in
Fig.~\ref{fig:reTScompact} at $t=2.1$, just after the source is turned on, shows
that the signal has instantaneously propagated around the closed lightlike line
in the $x$-direction. The right snapshot at $t=25$  shows that at late times the
wave has homogenized in the $y$-direction  as it decays into a long wavelength
mode of diminishing amplitude traveling in the $x$-direction.

\clearpage

\section{Marching algorithm for the half-plane problem}
\label{sec:alghalf}

We first consider the initial-boundary value problem for
\begin{equation}
      \partial_t (\partial_x \Phi+a \Phi)= \partial_x^2 \Phi
      \label{eq:mod}
\end{equation}
with the semi-discrete approximation
\begin{equation}
      \partial_t \bigg  (D_{0x}\Phi+a \Phi \bigg ) =D_{+x} D_{-x}  \Phi, 
      \label{eq:sdmod}
\end{equation}
in the half-plane $0\le x\le \infty$, with Dirichlet boundary conditions at the
inner time-like boundary $x=0$. The lines $t=\text{const}$ are outgoing
characteristics. For this problem, there is no periodicity in the $x$ direction
so that the discrete Fourier approach of Sec.~\ref{sec:fourlap} is no longer
appropriate for a numerical algorithm. Although a different spectral (or
pseudo-spectral) algorithm could be adopted, we pursue a local finite
difference approach. In this case the method of lines is not applicable because
reduction of (\ref{eq:sdmod}) to a first order system does not lead to a system
of ordinary differential equations in time.  Instead, we construct a marching
algorithm along the outgoing characteristics, as first developed  
in~\cite{march}. 

We discretize (\ref{eq:sdmod}) in both time and space on a grid with boundary at
$x=0$. This leads to the finite difference equation
\begin{equation}
    D_{0t} (D_{0x} \Phi+a \Phi) =D_{+x} D_{-x}  \Phi, 
    \label{eq:dmod}
\end{equation}
where $t_n=n\Delta t$. We denote $\Delta t=\lambda \Delta x  = \lambda h$, in
terms of the Courant factor $\lambda$. Setting $\Phi^n_j =\Phi(n\lambda h,jh)$
and centering (\ref{eq:dmod}) about the virtual grid point $(n+\frac{1}{2},
j-\frac{1}{2})$, we use the second order accurate approximations
\begin{equation}
      \partial_t D_{0} \Phi =\frac{ \Phi^{n+1}_{j} - \Phi^{n+1}_{j-1}- \Phi^{n}_{j} 
      +\Phi^{n}_{j-1}}{\lambda h^2},
      \label{eq:dtxapprox}
\end{equation}
\begin{equation}
      \partial_t \Phi =\frac{ \Phi^{n+1}_{j} +\Phi^{n+1}_{j-1}
           - \Phi^{n}_{j} -\Phi^{n}_{j-1}}{2\lambda h},
      \label{eq:dtapprox}
\end{equation}
\begin{equation}
      \partial_x \Phi =\frac{ \Phi^{n+1}_{j} -\Phi^{n+1}_{j-1}+ \Phi^{n}_{j} -\Phi^{n}_{j-1}}{2 h},
      \label{eq:dxapprox}
\end{equation}
and
\begin{equation}
 D_+D_-  \Phi =\frac{ \Phi^{n+1}_{j} - 2\Phi^{n+1}_{j-1}+\Phi^{n+1}_{j-2} 
                + \Phi^{n}_{j+1} - 2\Phi^{n}_{j} +\Phi^{n}_{j-1} }{2h^2}.
                \label{eq:lap}
\end{equation}
With these approximations (\ref{eq:dmod}) is equivalent to 
\begin{equation}
 \frac{ \Phi^{n+1}_{j} - \Phi^{n+1}_{j-1}- \Phi^{n}_{j} +\Phi^{n}_{j-1}}{\lambda h^2}
  + \frac{a( \Phi^{n+1}_{j} +\Phi^{n+1}_{j-1}- \Phi^{n}_{j} -\Phi^{n}_{j-1})}{2\lambda h}
  = \frac{ \Phi^{n+1}_{j} - 2\Phi^{n+1}_{j-1}+\Phi^{n+1}_{j-2} 
                + \Phi^{n}_{j+1} - 2\Phi^{n}_{j} +\Phi^{n}_{j-1} }{2h^2}
                \label{eq:ddmod}
\end{equation}
so that
\begin{eqnarray}
 \bigg (1+(ah/2)-(\lambda/2)  \bigg) \Phi^{n+1}_{j} &=&
      \bigg(1-(ah/2)-\lambda   \bigg) \Phi^{n+1}_{j-1}
   +(\lambda/2)   \Phi^{n+1}_{j-2}   + (\lambda/2) \Phi^{n}_{j+1}  \nonumber \\
  & +&\bigg (1+ (ah/2) - \lambda  \bigg )  \Phi^{n}_{j} 
      +\bigg (-1+ (ah/2) + (\lambda/2)  \bigg )  \Phi^{n}_{j-1} ,
      \label{eq:march}
\end{eqnarray} which determines $\Phi^{n+1}_{j}$, in terms of
$(\Phi^{n+1}_{j-1},\Phi^{n+1}_{j-2},\Phi^{n}_{j+1},\Phi^{n}_{j},\Phi^{n}_{j-1})$.
For $a=0$, this is identical to the approximation obtained in~\cite{march} which
was obtained from an integral identity satisfied by a scalar wave  at the
corners of a characteristic parallelogram. 

The marching algorithm based upon (\ref{eq:march}) proceeds as follows. Let
$\Phi^{n}_{j}$ be given on time level $t_n$ and let  $\Phi^{n+1}_{0}$ and 
$\Phi^{n+1}_{1}$ be given on time level $t_{n+1}$. Then (\ref{eq:march})
determines $\Phi^{n+1}_{2}$ and then, sequentially,  $\Phi^{n+1}_{3},
\Phi^{n+1}_{4}, \Phi^{n+1}_{5} \dots$. After updating the $t_{n+1}$ time level,
this outward march along the characteristic is repeated to update $t_{n+2}$. The
required initial data are $\Phi^{0}_j$, which are supplied by the characteristic
initial data $\Phi(0,x)$. The required boundary data are $\Phi^{n}_0$, which are
supplied by the Dirichlet boundary data $\Phi(t,0)$.  In addition, a start-up
algorithm is needed to obtain $\Phi^{n}_1$ (see below).

The stability of the algorithm can be determined by analyzing the behavior of
the Fourier-Laplace modes \begin{equation} \Phi^n_j= e^{st_n} e^{i\omega j h},
\quad t_n=n\lambda h . \end{equation} Substitution into (\ref{eq:ddmod}) leads,
after some algebra, to the  amplification factor
\begin{equation}
   \hat Q=e^{s\lambda h}=  \frac { 1-\lambda \sin^2(\omega h/2)
   +i\cot(\omega h/2) (\lambda \sin^2(\omega h/2) -(ah/2))  }
   {1-\lambda \sin^2(\omega h/2)
   -i\cot(\omega h/2) (\lambda \sin^2(\omega h/2) +(ah/2)) }.
   \label{eq:ampl}
 \end{equation} 
Thus, as found in~\cite{march},  $| \hat Q| =1$ for the case $a=0$, i.e. the
algorithm is unimodular. For $a>0$,  $| \hat Q| <1$ and the algorithm is damped.
The algorithm is unstable for $a<0$.

An algorithm is unconditionally stable if $|\hat Q | \le 1$ and stable if $|\hat
Q |$ has an upper bound independent of the values of $\omega$ and $h$. The above
algorithm is unconditionally stable for $a\ge 0$. In addition, it is convergent
provided $\lambda < 1$ so that the Courant-Friedrichs-Lewy (CFL) condition
(that the numerical domain of dependence contain the analytic domain of
dependence) is satisfied. However, the unimodular stability for the limiting
case $a=0$ makes the algorithm prone to instabilities when extended to higher
dimensional systems. In addition, see~\cite{luisdiss} for a discussion of
potential nonlinear instabilities and a technique to control them by introducing
artificial dissipation into (\ref{eq:ddmod}). Modulo such higher order terms,
the basic finite difference approximation (\ref{eq:ddmod}) appears to be the
unique 2-level algorithm which is stable, convergent and second order accurate,
i.e. we have not been able to find any other other algorithm with these
properties.

The generalization of the marching algorithm to higher dimensions is
straightforward using the techniques described in~\cite{march}. However,
instabilities can arise from the introduction of additional lower order terms. A
critical case is the 2-spatial dimension wave equation
\begin{equation}
   \partial_t (\partial_x \Phi +a\Phi )=\partial_x^2 \Phi  
   +\partial_y^2 \Phi- 2b \partial_y \Phi, \quad a>0.
   \label{eq:bwave}
 \end{equation}
The corresponding CFL condition is $\lambda < 1/2$, which is determined by the
characteristics in the  $(t,\pm y)$ planes. As before, we discretize the
$y$-dependence  on a periodic grid and write $\Phi^{n}_{j_1,j_2}=
\Phi(t_n,x_{j_1},y_{j_2})$. We introduce the second order approximation
\begin{equation}
   [\partial_y^2 \Phi- 2b \partial_y \Phi]^{n+1/2}_{j_1-1/2,j_2} =
           \frac{1}{2}  (D_{+y}D_{-y}-2bD_{0y})(\Phi^{n+1}_{j_1-1,j_2}+ \Phi^{n}_{j_1.j_2} ),
           \label{eq:yderiv}
\end{equation}
which does not involve the values $\Phi^{n+1}_{j_1,j_2}$ so that the marching
algorithm can be extended to 2-dimension. In order to analyze the stability of
the resulting marching algorithm, we consider the modes
\begin{equation}
     \Phi^n_{j_1,j_2}= e^{sn\lambda h} e^{i\omega_1 j_1 h} e^{i\omega_2 j_2 h}.
\end{equation}

From a calculation analogous to that leading to (\ref{eq:ampl}), we find
\begin{equation}
    |\hat Q|^2= \frac{ \bigg ( 1-A -hB \cot(h\omega_1/2)\bigg )^2
          + \bigg ((A- (ah/2) )\cot(\omega_1 h/2)  -hB    \bigg )^2}
          {\bigg  (  1-A +hB \cot (\omega_1 h/2) \bigg )^2 
          +  \bigg  ((A+(ah/2)) \cot(\omega_1 h/2) +hB    \bigg )^2 } ,
          \label{eq:hatqcot}
\end{equation}
where
\begin{equation}
     A=\lambda ( \sin^2(h\omega_1/2)+\sin^2(h\omega_2/2)) 
 \end{equation}
and
\begin{equation}
    B= \frac{\lambda  b\sin(\omega_2 h)}{2} .
\end{equation}
Thus 
\begin{equation}
    |\hat Q|^2=1+ \frac{ - 4hB \cot(h\omega_1/2)
           -2ah A \cot^2(h\omega_1/2 )    }
          { \bigg (1-A +hB \cot(h\omega_1/2)\bigg )^2 
          + \bigg  ((A  +(ah/2) ) \cot(h\omega_1/2) +hB\bigg  )^2    } .
          \label{eq:cot}
\end{equation}
For $b=0$ and $a\ge 0$, $|\hat Q|\le 1$ and the algorithm is unconditionally
stable. 

In order to analyze the effect of $b \ne 0$, first
consider the continuum approximation
$h \approx 0$ with  bounded values of $(\omega_1,\omega_2)$, for which
\begin{equation}
  \bigg (- 4hB \cot(h\omega_1/2) -2ah A \cot^2(h\omega_1/2) \bigg )\sin^2(h\omega_1/2)
      \approx -\frac{\lambda h^3}{2}\bigg (2b \omega_1 \omega_2
      +a( \omega_1^2+ \omega_2^2)\bigg )
      \le \frac{\lambda h^3}{2}(|b| -a)( \omega_1^2+ \omega_2^2),
\end{equation}
so that, referring to (\ref{eq:hatqcot}), $|\hat Q|\le 1$  for $|b|\le a$ and
the algorithm remains unconditionally stable. More generally in this
approximation,
\begin{eqnarray}
    |\hat Q|^2 &\approx& 1 - \frac{2\lambda h \bigg (2b \omega_1 \omega_2
      +a( \omega_1^2+ \omega_2^2)\bigg )}{\omega_1^2+a^2} \nonumber \\
      &\le& 1 - \frac { 2\lambda h \bigg (a(\omega_2+{b\over a} \omega_1)^2
      -{b^2\over a} \omega_1^2 \bigg ) } {\omega_1^2+a}  
       \le  1 + \frac {2\lambda h b^2} {a}, \nonumber 
\end{eqnarray}
so that the growth rate is bounded independent of $(\omega_1,\omega_2)$. Hence
the algorithm remains numerically stable but reflects the existence of
exponentially growing modes in the analytic problem. In this approximation, note
that for $a=0$ and $b\ne 0$ that
\begin{equation}
    |\hat Q|^2 \approx 1 - 4\lambda b{ \omega_2 \over  \omega_1}, 
\end{equation}      
so there would be a catastrophic instability for $(-b\omega_2 )/ \omega_1
\rightarrow \infty$. This again reflects the important role of the condition
$a>0$ in determining the well-posedness of the problem.

Next, for $a>0$, consider the limiting cases where either $\omega_1 \rightarrow
\infty$ or $\omega_2 \rightarrow \infty$, or both, as $h\rightarrow 0$. We
express (\ref{eq:cot}) in the form
\begin{equation}
    |\hat Q|^2=1+ \frac{-2h \bigg(B(Aa)^{-1/2} +(Aa)^{1/2}\cot(h\omega_1/2) \bigg )^2
           +2hB^2  (Aa)^{-1}}
         { \bigg (1-A+hB \cot(h\omega_1/2)\bigg )^2 
          +  \bigg  ((A+(ah/2))\cot(h\omega_1/2)+hB \bigg  )^2    }  ,
\end{equation}
so that
\begin{eqnarray}
    |\hat Q|^2 &\le& 1+ \frac{  2hB^2  (Aa)^{-1} }
          { \bigg (1-A+hB \cot(h\omega_1/2)\bigg )^2 
          +  \bigg  ((A+(ah/2))\cot(h\omega_1/2)+hB \bigg  )^2    }\nonumber \\
         & \le& 1+ \frac{  2hb^2/a }
          { \bigg (1-A+hB \cot(h\omega_1/2)\bigg )^2 
          +  \bigg  ((A+(ah/2))\cot(h\omega_1/2)+hB \bigg  )^2    } ,
          \label{eq:bstable}
\end{eqnarray}
where we have used the identity $\sin^4 \theta = 4 \sin^2 (\theta/2)\cos^2
(\theta/2)$ to obtain
\begin{equation}
  B^2  A^{-1} =\frac {b^2\sin^2(h\omega_2/2 )\cos^2(h\omega_2/2 )}
       {\sin^2(h\omega_1/2 )+\sin^2(h\omega_2/2 )} \le b^2.
\end{equation}

We observe that uncontrolled growth can only occur if there are values of
$(\omega_1,\omega_2)$ such that the denominator in (\ref{eq:bstable}) vanishes,
which requires
\begin{equation}
        hB=-\frac{1-A}{\cot(h\omega_1/2)}=-(A+\frac{ah}{2})\cot(h\omega_1/2).
        \label{eq:qstable}
\end{equation}
The second equality in (\ref{eq:qstable}) implies
\begin{equation}
       \cot^2(h\omega_1/2)=\frac{1-A}{A+(ah/2)} 
        =\frac{1-\lambda\bigg (\sin^2(h\omega_1/2 )+\sin^2(h\omega_2/2 ) \bigg )}
        {\lambda\bigg (\sin^2(h\omega_1/2 )+\sin^2(h\omega_2/2 )\bigg )+(ah/2)} ,
\end{equation}
so that solving for $\sin^2(h\omega_1/2 )$ gives
\begin{equation}
      \sin^2(h\omega_1/2 )
        =\frac{\lambda\sin^2(h\omega_2/2 ) +(ah/2)}
        {1-  \lambda+(ah/2)} 
        \label{eq:sin}
\end{equation}
as a necessary condition for an unstable mode. However, because $B=O(h)$, a
straightforward analysis shows that the first equality in (\ref{eq:qstable})
rules out such modes for reasonably small values of $h$.  As an example,
consider the mode  $\sin^2(h\omega_1/2)=\sin^2(h\omega_2/2)=1/2$, 
(\ref{eq:sin}) with $\lambda =(1/2) -(ah/4)$, which satisfies (\ref{eq:sin}) and
is consistent with the Courant condition $\lambda<1/2$.  Then the first equality
in (\ref{eq:qstable}) would require,
\begin{equation}
         |b|h = 2|1-\lambda^{-1} |,
\end{equation}
which would not be satisfied. The robust stability boundary test described in
Sec.~\ref{sec:stability} verifies this conclusion that the marching algorithm is
stable.

In Sec.~\ref{sec:testbound},  we present tests and simulations of the
null-timelike problem
\begin{equation}
   \partial_t (\partial_x \Phi +a\Phi )=\partial_x\bigg ((1-x)^2\partial_x \Phi  \bigg )
   +\partial_y^2 \Phi -2b \partial_y \Phi , \quad 0\le x \le 1, \quad 0\le y <1, \quad t \ge 0,
   \label{eq:swave}
\end{equation}  
with periodicity in $y$. We assign Dirichlet boundary values at $x=0$. The
boundary at $x=1$ is an ingoing characteristic surface and requires no boundary
condition.  It is analogous to the boundary at null infinity obtained by conformal
compactification. The finite difference approximation (\ref{eq:lap}) is modified
according to  
\begin{equation}
\partial_x\bigg ((1-x)^2\partial_x \Phi  \bigg )=
  (1-x)^2\partial^2_x \Phi -2(1-x) \partial_x \Phi
 \approx (1-x)^2 D_{+x}D_{-x}  \Phi -2(1-x)  D_{0x} \Phi ,
\end{equation}
which is evaluated at the mid-point $(j_1-1/2,j_2)$ at  time  $t_{n+1/2}$ by
\begin{equation}
 (1-x)^2 D_{+x}D_{-x}  \Phi =
   \frac{1}{2} (1-x_{j_1-1})^2 D_{+x}D_{-x}  \Phi^{n+1}_{j_1-1,j_2}
   +   \frac{1}{2} (1-x_{j_1})^2 D_{+x}D_{-x}  \Phi^{n}_{j_1,j_2}
\label{eq:dxsq}
\end{equation}
and
\begin{equation}
    -2(1-x)  D_{0x} \Phi = - (1-x_{j_1-1/2}) D_{0x} \bigg ( \Phi^{n+1}_{j_1-1/2,j_2} 
       +  \Phi^{n}_{j_1-1/2,j_2} \bigg ) .
\end{equation}

These approximations, along with the previous second order approximations for
the other terms in the wave equation, allow the values of $\Phi^{n+1}_{N_1,j_2}
$  at the outer (characteristic) boundary $x=1$ to be updated in terms of
values previously determined by the marching algorithm. The startup algorithm at
the inner boundary $x=0$ is more complicated since an approximation for
$\partial_x^2 \Phi$ requires a minimum of three grid points. In order to
circumvent this difficulty we write
\begin{equation}
  \partial_x^2 \Phi =\lambda \partial_x \partial_t \Phi
       -\partial_x (\lambda\partial_t -\partial_x)\Phi.
\label{eq:startD2x}
\end{equation} 
Here the first term on the right hand side only requires a two-point stencil as
in (\ref{eq:dtxapprox}). In addition, the second term can be approximated to
second order by a stencil involving only the first two points on the upper time level,
\begin{equation}
   \partial_x ( \lambda \partial_t-\partial_x ) \Phi^{n+1/2}_{1/2,j_2} = 
       \frac{1}{h}  D_{0x}(\Phi^{n+1}_{1/2,j_2}-\Phi^{n}_{3/2,j_2}) =
        \frac{1}{h^2} (\Phi^{n+1}_{1,j_2}-\Phi^{n+1}_{0,j_2}
        -\Phi^{n}_{2,j_2} +\Phi^{n}_{1,j_2}).
\label{eq:startDx}
\end{equation}
Along with the other terms in the wave equation,
this can be used to determine $\Phi^{n+1}_{1,j_2} $ 
in terms of the Dirichlet boundary data  $\Phi^{n+1}_{0,j_2}$ 
and previously determined values.

\section{Simulation of the null-timelike strip problem}
\label{sec:testbound} 

Here we apply the marching algorithm\ described in Sec.~\ref{sec:alghalf} to
simulate the initial-boundary value problem for the wave equation (\ref{eq:swave}),
\begin{equation}
   \partial_t (\partial_x \Phi +a\Phi )=\partial_x\bigg ((1-x)^2\partial_x \Phi  \bigg )
   +\partial_y^2 \Phi -2b \partial_y \Phi ,  \quad t \ge 0 ,
   \label{eq:swave2}
\end{equation}  
in the strip $0\le x \le 1$,  $ 0\le y <1$, with periodicity in the $y$-direction.
We assign initial data and Dirichlet boundary data
\begin{equation}
    \Phi(0,x,y)=f(x,y) \quad  \Phi (t,0,y) =q(t,y).
   \label{eq:halfpl}
\end{equation}
The principle part of the wave operator in (\ref{eq:swave2}) corresponds to the
metric 
$$
    ds^2= -4(1-x)^2 dt^2 - 4dt dx +dy^2.
$$
This implies that the boundary at
$x=1$ is an ingoing characteristic hypersurface so that no boundary condition should
be applied there.

This strip problem with $a>0$ was shown to be well posed in~\cite{wpsw},
where an energy estimate
established that the solution and its derivatives remain bounded.
We demonstrate this by the stability and convergence tests in Sec's.~\ref{sec:stability}
and \ref{sec:scon}.
The simulations in Sec's~\ref{sec:atests} illustrate the effects
of the lower order $a$-term.

\subsection {Stability}
\label{sec:stability}

\begin{figure}[ht]
\centering
\includegraphics*[width=10cm]{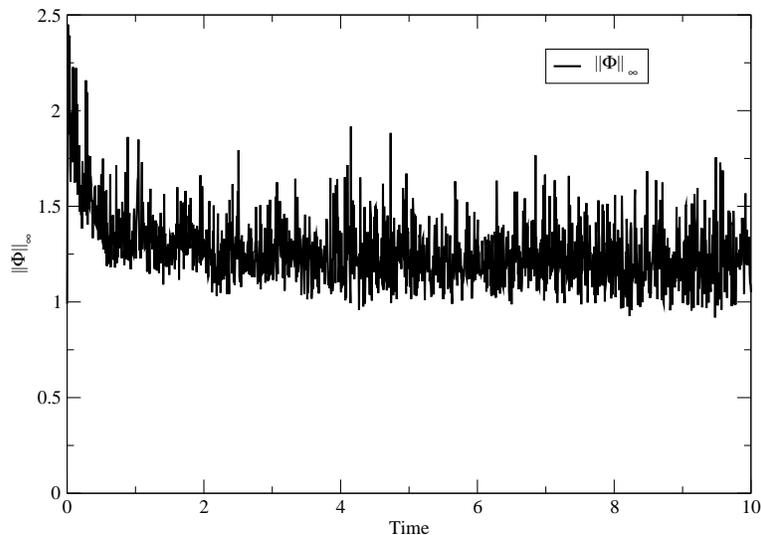}
\caption{ Time dependence of the $L_{\infty}$ norm $||\Phi||_\infty$ for 
the robust stability boundary test with $a=b=1$.}
\label{fig:RandLinf}
\end{figure}

The analysis in  Sec.~\ref{sec:alghalf} showed that the marching algorithm
for the problem  (\ref{eq:swave2}) is stable for $a>0$. For $a <0$, the problem
is ill posed and numerical stability cannot be expected. We confirm these
results by carrying out robust stability boundary tests~\cite{robust}
based upon random initial and boundary data. The test
is implemented using a random number generator that changes at each time-step. 
The test was run on a grid of size $N\times N=64^2$, with
$h=\Delta x =1/N$ and timestep $\Delta t = h/5$,  
with an amplitude for the random noise of $A=1$, for a time $t=10$.
The graph of the time dependence of the  $L_{\infty}$ norm for the case
$a=b=1$ is shown in  Figure~\ref{fig:RandLinf}.
There is no sign of numerical instability.

The instability for cases $a<0$ is not as pronounced as for the
tests of the Cauchy problem. This is because the unstable modes
are propagated off the grid through the characteristic boundary at $x=1$.
Consequently, the robust stability test is not as effective as in the
case of the strip problem between two timelike boundaries. Nevertheless,
rapid unstable growth is evident for sufficiently negative values of $a$.

\subsection {Convergence}
\label{sec:scon}

\begin{figure}[ht]
\centering
\includegraphics*[width=8cm]{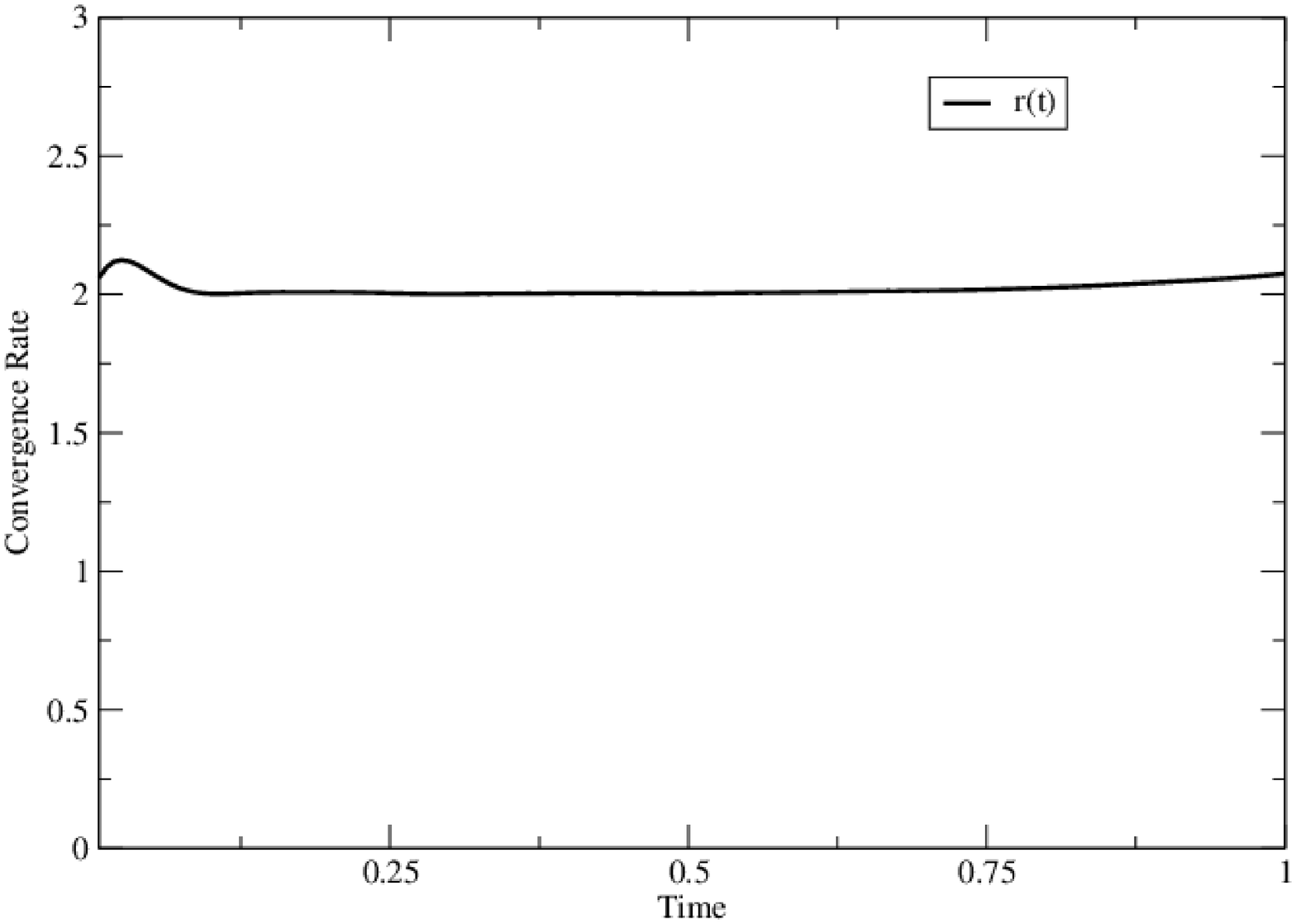}
\includegraphics*[width=8cm]{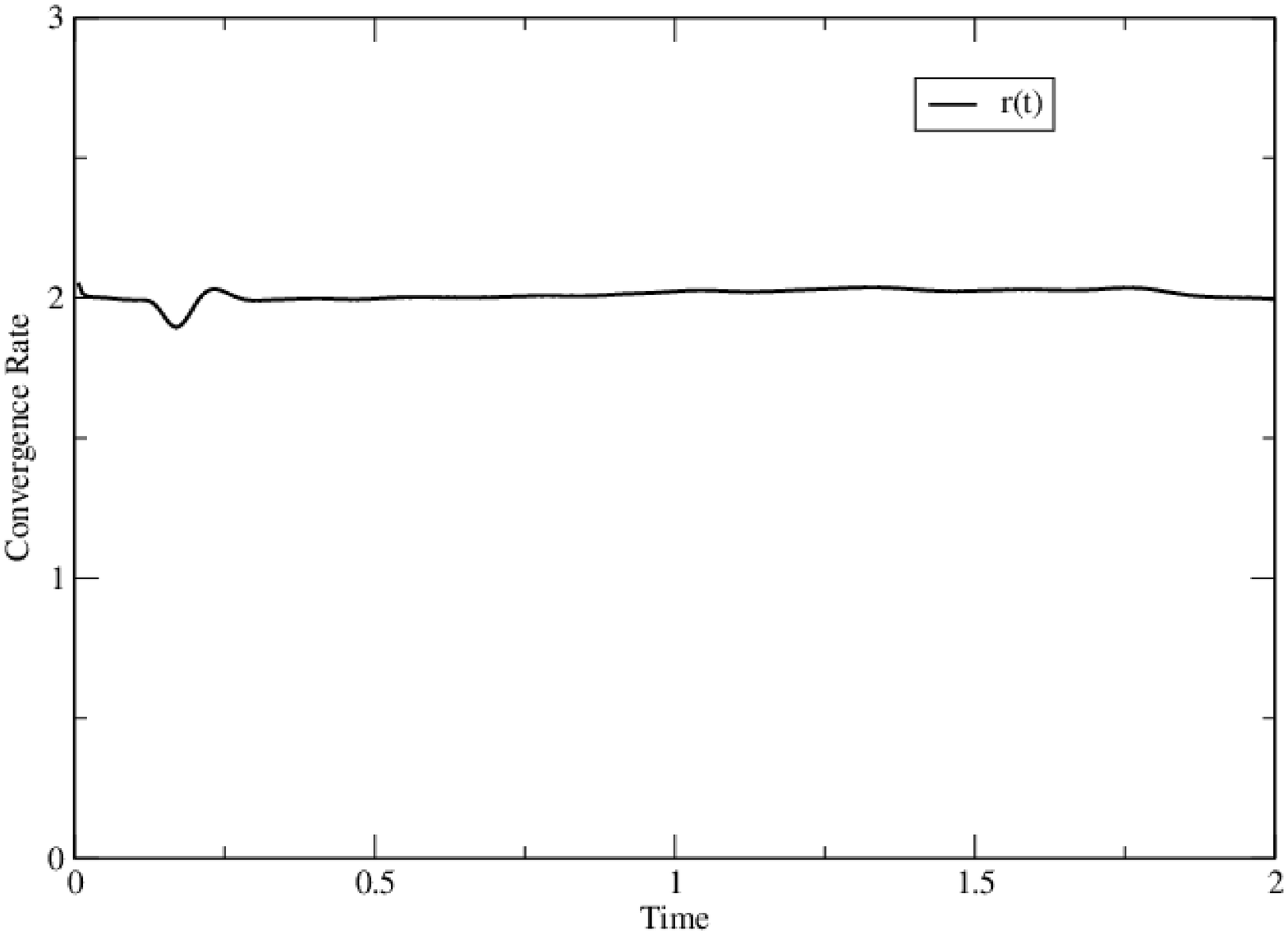}
\caption{
Left: Convergence rate $r(t)$ to the exact solution (\ref{eq:exact2D}) in the
interval $0\le t \le 1$, with $a=10$ and $b=0$. \\
Right: Cauchy convergence rate  $r(t)$ for simulation of the initial
pulse (\ref{eq:idata2D}) with  $a=b=1$.
}
\label{fig:ConvPlots}
\end{figure}

We carry out two convergence tests. In the first, the initial-boundary data
is based upon the exact
solution for $b=0$,
\begin{equation}
   \Phi =e^{st}  \phi(x)  \cos(\omega y), \quad s=-\frac{\omega^2}{a}
   \label{eq:exact2D}
\end{equation}
where 
\begin{equation}
       \phi(x)=e^{ sx/(1-x) } .
\end{equation}
In this test we set $\omega =2\pi$ and $a=10$ so that the exponential decay
of the solution can be resolved with reasonable sized grids
$N_{A}\times N_{A}$, with $N_A=64/A$ in the ratio $A=(1,2)$.
On these grids, the convergence to the exact solution is measured
with Courant factor $\Delta t_A / h_A =1/5$ in the interval $0\le t \le 1$. 
The convergence rate $r(t)\approx2$, shown in the
left plot  of Fig.~\ref{fig:ConvPlots},  is in excellent accord
with the second order finite difference approximations.

We also test Cauchy convergence for the case $a=b=1$
for the simulation of  the initial  pulse 
\begin{eqnarray}
\label{eq:compactP}
\Phi(0,x,y) &=& A \left [ (x-x_1)(y-y_1)(x-x_2)(y-y_2)\right ]^4 , 
              \quad  x_1\le x \le x_2,  \, y_1\le y \le y_2 ,\nonumber \\
\Phi(0,x,y) &=& 0, \quad  {\rm otherwise},  
	\label{eq:idata2D}       
\end{eqnarray}
with vanishing boundary data.
We take $x_1=y_1=0.1$, $x_2=y_2=0.9$ and amplitude $A=10^{5}$. 
The Cauchy convergence rate $r(t)$, measured for three grids of 
size $N_{A}\times N_{A}$, with $N_A=128/A$, with $A=(1,2,4)$,and 
Courant factor $\Delta t_A / h_A =1/5$, is shown in
the right plot of Fig.~\ref{fig:ConvPlots}. 
We again obtain clean second order convergence $r(t)\approx2$.

\subsection {Damping, radiation tails and the $a$-term}
\label{sec:atests}

\begin{figure}[ht]
\centering
\includegraphics*[width=8cm]{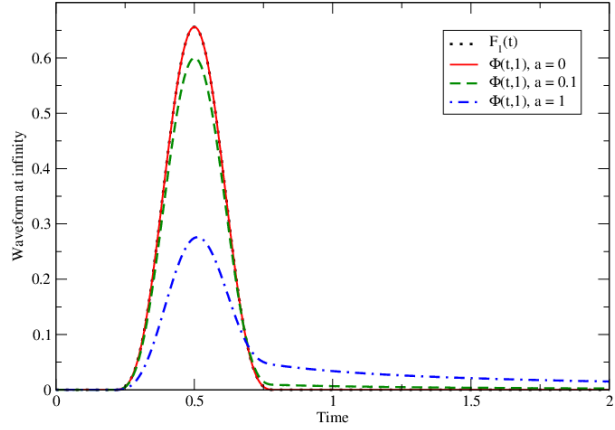}
\caption{The waveform at the outer characteristic boundary
corresponding to the purely outgoing data (\ref{eq:pdata}).
The $a$-term produces a tail which decays
on a time scale increasing with $a$.}
\label{fig:PulseDampinf}
\end{figure}

In order to reveal the effects of the $a$-term in (\ref{eq:swave2}) first
consider the exact 1-dimensional solutions for the case $a=0$,
\begin{equation}
     \Phi(t,x) =F_1(t)+ F_2(t+\frac{1}{1-x}).
     \label{eq:exact}
\end{equation}     
Here $F_1(t)$, with $F_2=0$, describes a purely outgoing wave  emanating from
the inner boundary. For the simulation of such a purely outgoing wave, we choose
the pulsed signal
$$  F_1(t) = A[(t-t_1)(t-t_2)]^4, \quad  t_1=0.2 ,\,  t_2 =0.8, \quad F_1(t) =0 \,
\text{for}\, t<t_1\, \text{and} \,t>t_2,
$$
corresponding to the initial and boundary data
\begin{equation}
         \Phi(0,x)=0 \, , \quad  \Phi (t,0) = F_1(t).
         \label{eq:pdata}
\end{equation}
We choose the amplitude $A=10^4$ to make the peak of order unity. For $a=0$ the
profile $\Phi(t,1)$ of the waveform passing through the outer boundary is
identical with the inner boundary data, i.e. $\Phi(t,1) =F_1(t)$, and the error
is the order of machine precision. For $a>0$. the solution to (\ref{eq:swave2})
resulting from the data (\ref{eq:pdata}) is no longer given by (\ref{eq:exact}) and
shows the effects of backscatter. Figure~\ref{fig:PulseDampinf}, which compares the
waveforms $\Phi(t,1)$ resulting from the initial data (\ref{eq:pdata}) for the
cases $a=0$, $a=0.1$ and $a=1$, illustrates the damping of the outgoing waveform
due to the $a$-term. The $a$-term produces a tail to the waveform which decays
on a time scale increasing with the size of $a$.

\begin{figure}[ht]
\centering
\includegraphics*[width=8cm]{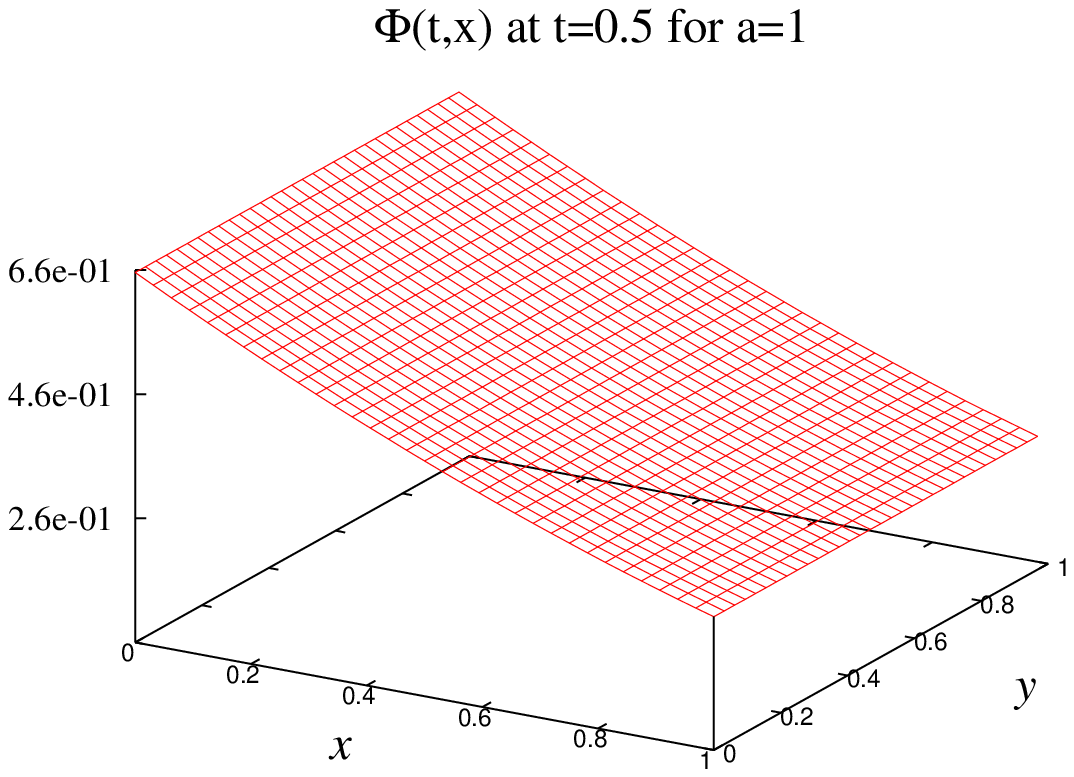}
\includegraphics*[width=8cm]{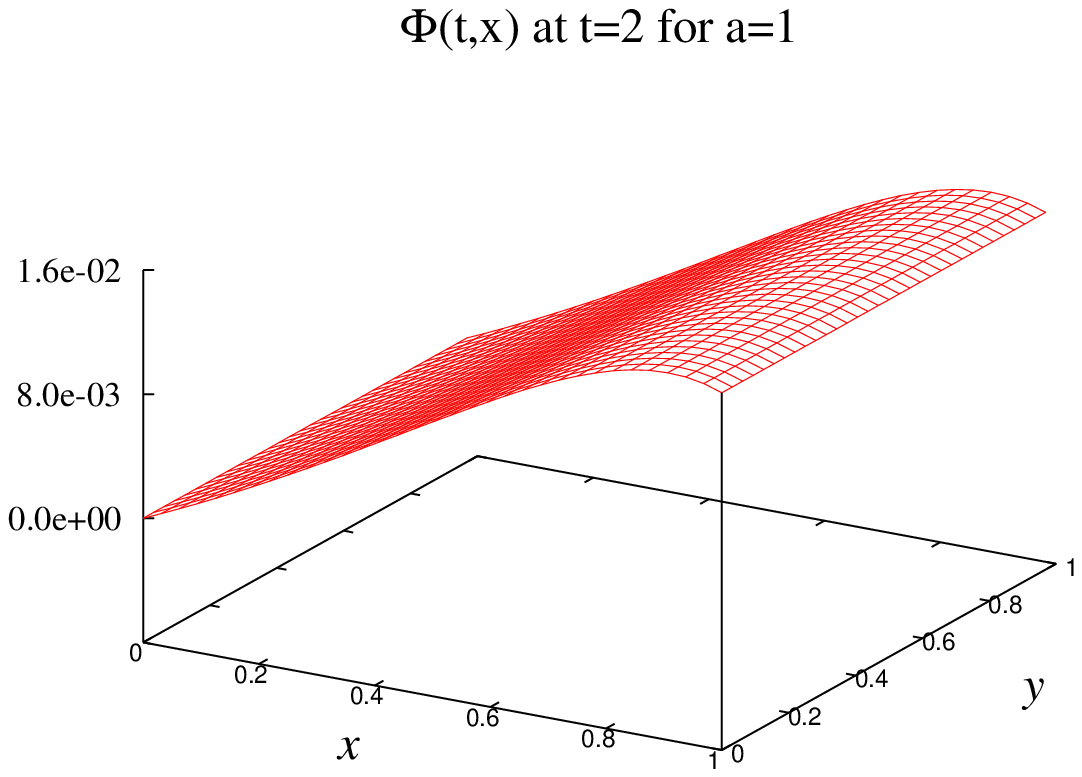}
\caption{Snapshots $\Phi(t,x)$ for  $a=1$, with boundary data  (\ref{eq:pdata}). 
The left plot  in the middle of the signal  at $t=0.5$ is dominated by
 the negative slope at the outer boundary resulting from (\ref{eq:negslope}).
The right plot at $t=2$ shows that this slope becomes small during the tail decay.}
\label{fig:PulseA1}
\end{figure}

Snapshots of $\Phi(t,x)$ for  $a=1$ are given in Fig.~\ref{fig:PulseA1}  at
times $t=0.5$ (in the middle of the signal $F_1(t)$) and $t=2.0$
(after the signal has turned off). The snapshots exhibit a distinctive, almost
horizontal slope near the outer boundary, which can be explained by
evaluating (\ref{eq:swave2}) at $x=1$, 
\begin{equation}
           \partial_t \bigg ( \partial_x \Phi(t,1) -a \Phi(t,1) \bigg ) =0,  \, \quad a=1,
 \end{equation}  
which implies 
\begin{equation}
              \partial_x \Phi(t,1) =-\Phi(t,1),
              \label{eq:negslope}
 \end{equation}  
since $\Phi(0,x) =0$. 
During the signal, when $\Phi(t,1)$ is large and positive, this results
in a negative slope at $x=1$ but during the tail, as $\Phi(t,1)$ decays to zero,
this slope becomes very small. 

\begin{figure}[ht]
\centering
\includegraphics*[width=8cm]{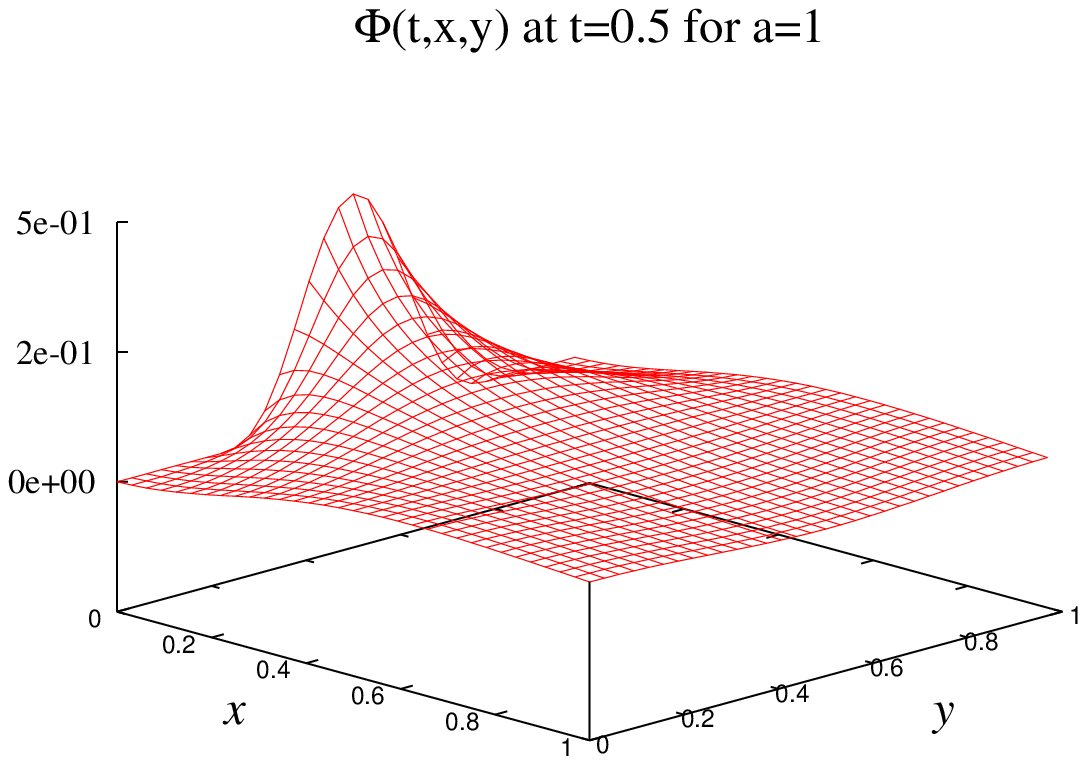}
\includegraphics*[width=8cm]{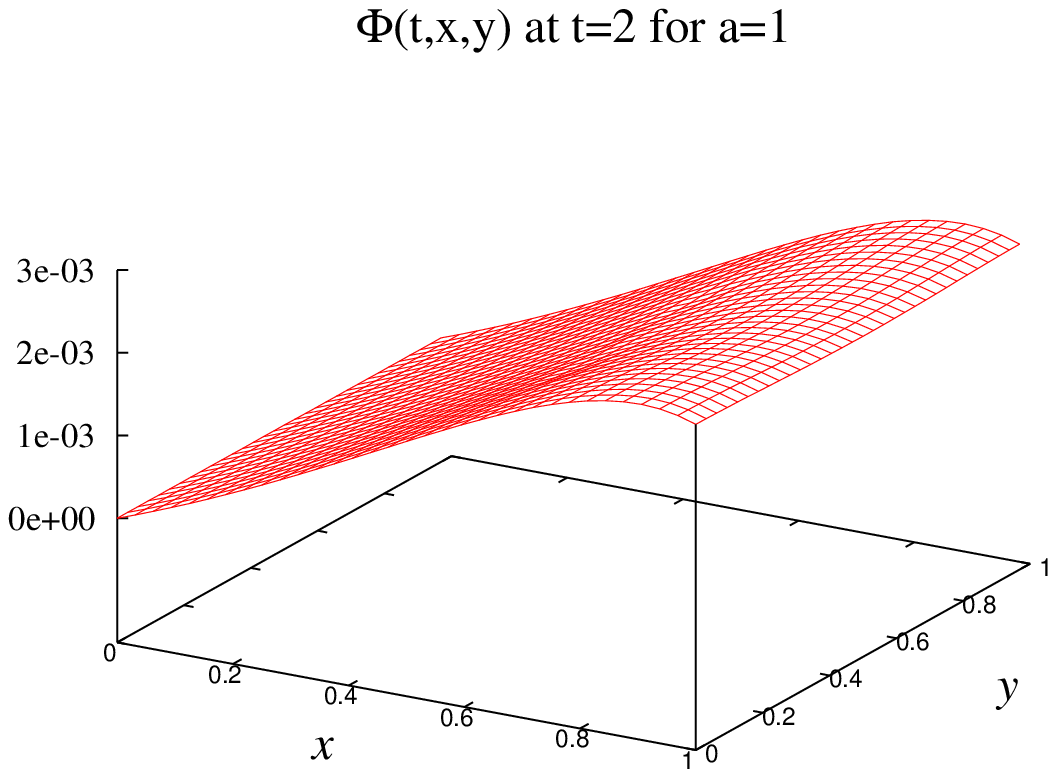}
\caption{Snapshots $\Phi(t,x,y)$ for  $a=1$ and $b=0$, with pulse shaped
boundary data (\ref{eq:pydata}) during the middle of the
signal at $t=0.5$ (left plot) and during the tail decay
at $t=2$ (right plot). The similarity to the 1-dimensional case
in Fig.~\ref{fig:PulseA1} is evident.}
\label{fig:PulseA1Y}
\end{figure}

These 1-dimensional results shed light on the behavior in the 2-dimensional
case obtained by adding a compact $y$-dependence to the pulsed
boundary data,
\begin{equation}
         \Phi(0,x,y)=0\, , \quad  \Phi (t,0,y) =A[(t-t_1)(t-t_2)(y-y_1)(y-y_2)]^4, 
\quad  (t_1,y_1)=0.2 ,\,  (t_2,y_2) =0.8.
         \label{eq:pydata}
\end{equation}
The left plot of Fig.~\ref{fig:PulseA1Y} shows a snapshot of the wave in the
middle of the signal at $t=0.5$. Where the $y$-dependence of the pulse is
peaked, it is similar to the left plot  Fig.~\ref{fig:PulseA1} for the
1-dimensional case. Similarly the snapshot during the tail decay at $t=2$ in the
right plot of Fig.~\ref{fig:PulseA1Y}
is almost identical to the right plot of Fig.~\ref{fig:PulseA1}, differing only by
some backscatter of the original pulse.

\clearpage

\section{Summary}

Analytic results regarding the importance of the $a$-term
for the well-posedness of the
characteristic initial-boundary value problems for the wave equation have been
extended to the finite difference formulation. Based upon these results, we have
developed computational evolution algorithms for the scalar wave equation in
characteristic coordinates. We proved the numerical stability of the algorithms
analytically, by means of both Fourier-Laplace theory  and the energy method.
The approach allowed us to individually test the finite difference code for the
pure Cauchy problem and for the initial-boundary value problem in a strip.
The main result for both problems is that numerical stability is controlled by
the condition $a>0$, an
important feature which had been overlooked in treatments of the characteristic
initial value problem for the wave equation. 

The pure Cauchy problem was implemented with periodic boundary conditions so
that characteristics formed closed timelike curves. This gave rise to simulations
in which a signal propagated instantaneously back to the source. The evolution
code for the strip problem, with timelike inner boundary and characteristic
outer boundary, was modeled upon the characteristic marching
algorithm, which has been used for characteristic evolution in general
relativity. The knowledge gained from the model problems considered here
should be of benefit to a better understanding of the gravitational case and other
applications of characteristic evolution~\cite{lr}.

\begin{acknowledgments}

The research was supported by NSF grants
PHY-0854623 and PHY-1201276 to the University of Pittsburgh
and by NSF grant PHY-0969709 to Marshall University.

\end{acknowledgments}

\appendix

\end {document}